\documentclass{aa}

\newcommand{\degree}{^\circ}

\usepackage{placeins}

\usepackage{subfigure}

\usepackage{siunitx}
\usepackage{graphicx}
\usepackage[separate-uncertainty=true]{siunitx}

\def\apjl{\rmfamily{ApJ~}}        % Astrophysical Journal, Letters
\def\apjs{\rmfamily{ApJS~}}       % Astrophysical Journal, Supplement

\def\aap{\rmfamily{A\&A~}}        % Astronomy and Astrophysics

\def\ao{\rmfamily{Appl.~Opt.~}}   % Applied Optics

\usepackage{txfonts}
\usepackage{hyperref}
\hypersetup{
 colorlinks=true,
 citecolor=blue,
 linkcolor=blue,
 urlcolor=blue}

\usepackage{xcolor}

\begin{document}

   \title{Multi-scale radiative-transfer model of the protoplanetary disc DoAr 44}

   \author{M. Souza de Joode\inst{1,2}
          \and
          M. Bro\v z\inst{1}
          }

   \institute{
   \inst{1}Charles University, Faculty of Mathematics and Physics, Astronomical Institute, V Hole\v{s}ovi\v{c}k\'{a}ch
 2, 18000 Prague, Czech Republic \\
   \inst{2}Heidelberg University, Department of Physics and Astronomy, Im Neuenheimer Feld 226, 69120 Heidelberg, Germany  \\ \email{marco.souza\_de\_joode@stud.uni-heidelberg.de}
             }

   \date{Received December 17, 2024; accepted July 18, 2025}

\abstract
{} 
{We aim to construct a comprehensive global multi-scale kinematic equilibrium radiative-transfer model for the pre-transitional disc of DoAr\,44 (Haro 1-16, V2062 Oph) in the Ophiuchus star-forming region. This model integrates diverse observational datasets to describe the system, spanning from the accretion region to the outer disc.} 
{Our analysis utilises a large set of observational data, including ALMA continuum complex visibilities, VLTI/GRAVITY continuum squared visibilities, closure phases, and triple products, as well as VLT/UVES and VLT/X-shooter H\(\alpha\) spectra. Additionally, we incorporated absolute flux measurements from ground-based optical observatories, Spitzer, IRAS, the Submillimeter Array, the IRAM, or the ATCA radio telescopes. These data sets were used to constrain the structure and kinematics of the object through radiative-transfer modelling.} 
{Our model reveals that the spectral line profiles are best explained by an optically thin spherical inflow/outflow within the co-rotation radius of the star, exhibiting velocities exceeding 380\,km/s. The VLTI near-infrared interferometric observations are consistent with an inner disc extending from 0.1 to 0.2\,au. The ALMA sub-millimetre observations indicate a dust ring located between 36 and 56\,au, probably related to the CO$_2$ condensation line. The global density and temperature profiles derived from our model provide insight into an intermediate disc, located in the terrestrial planet-forming zone, which has not yet been spatially resolved.} 
{}

   \keywords{Protoplanetary discs -- Accretion, accretion discs -- Submilimeter: planetary systems -- Radiative transfer -- Techniques: interferometric}

   \maketitle
   
\section{Introduction}

   Protoplanetary discs are reservoirs of gas and dust surrounding young stellar objects and the birthplace of planets. Understanding them is necessary to understand the formation of exoplanetary systems and our own solar system. Various disc morphologies have been suggested based on the broad band spectral energy distribution, and now are being directly imaged either via sub-mm-observations or adaptive optics \citep{andrews2018disk, avenhaus2018disks}. Presumably, the youngest of these discs have monotonically decreasing density profiles, with a large inner cavity opening later, most probably due to the influence of forming ice-giants, creating so-called transitional discs \citep{van2023transition}, before dispersing a few millions of years later \citep{li2016lifetimes}. However, matter is still flowing through these cavities, as evidenced by the high accretion rates \citep{van2023transition}.

An intermediate phase has been identified \citep{epsaillat2007}, called the pre-transitional phase, where the young star is surrounded by an inner disc, encircled by a large, optically thin cavity, followed by an extended dust ring exhibiting mm-emission.

The object DoAr\,44 (Haro 1--16, V2062 Oph), located in the Ophiuchus star formation region, is one of these pre-transitional discs with a high water vapour luminosity in the inner disc \citep{salyk2015detection}, 40 times more than the structurally very similar disc PDS 70 \citep{perotti2023water}, where planet formation is confirmed by direct imaging by the Hubble Space Telescope (HST) \citep{zhou2021hubble}, ALMA \citep{benisty2021circumplanetary} and VLT/SPHERE \citep{keppler2018discovery}. DoAr\,44  has been investigated by VLTI/GRAVITY \citep{bouvier2020probing}, revealing a disc-like structure in the 0.2--1.5\,au region, VLT/SPHERE \citep{arce2023radio}, showing the disc extending beyond 15 au, and also ALMA Bands 3, 4, 6, 7 and 8 at resolutions up to \(\approx 25\) mas in Band 6 and 7 \citep{cieza2021ophiuchus, arce2023radio}, in both continuum and spectral lines, investigating the region from 5--80\,au, revealing a narrow dust ring with a brim around 47\,au.

A series of other works were devoted to this pre-transitional disc. For example, \citet{ricci2010dust} performed a 3.3\,mm continuum survey of the \(\varrho\)-Ophiuchi region, including DoAr\,44, using the Australian Telescope Compact Array (ATCA), measuring the spectral indices between the 1.3\,mm and 3.3\,mm fluxes. They concluded that the spectral indices of Ophiuchus sources are not statistically different from the sources in the Taurus/Auriga star-forming region, which they surveyed using the Plateau de Bure Interferometer (PdBI) in a previous study \citep{ricci2010dust2}, but differ significantly from the interstellar medium.

\citet{2010ApJ...723.1241A} then performed an interferometric survey using the Submillimeter Array (SMA) at 40\,au resolution in the Ophiuchus star-forming region. They tried fitting the surface density profile using the common \citet{1974MNRAS.168..603L} prescription:
\begin{equation}
    \Sigma = \Sigma_0 \left(\frac{R}{R_c}\right)^\gamma \exp \left[ -\left(\frac{R}{R_c}\right)^{2-\gamma} \right]\,,
\end{equation}
where \(\Sigma_0,R_c, \gamma\) are parameters. However, for several discs they found a poor fit, which could be explained by the fact that these discs (DoAr\,44, SR\,21, SR\,24 and WSB\,60) had a large central cavity. This was the first \textit{spatially} resolved observation of the cavities predicted by SED modelling. They continued their investigation of transitional discs with an additional SMA study  \citep{2011ApJ...732...42A}, finding that the dust in the cavity has been depleted by 4--6 orders of magnitude with respect to the \citet{1974MNRAS.168..603L} profile.

\citet{espaillat2010unveiling} modelled Infrared Telescope Facility (IRTF)  2.1–5.0 micron spectra of DoAr\,44 with \(R = 1500\) resolving power, and also Spitzer spectra, using analytical temperature prescriptions. They derived a higher accretion rate of \(\dot{M} = 9.3\times10^{-9} {\,M_\odot\,}{\rm yr^{-1}}\).

\citet{bouvier2020probing}  obtained K-band (2 -- 2.4\,\si{\micro \meter}) continuum and Br\(\gamma\) line visibilites of DoAr\,44 from VLTI/GRAVITY. They concluded that the Br\(\gamma\) emission is located inward of the continuum emission, suggesting that the optically thin gas flows, where the line emission is created, are located within the truncation radius. They used the \citet{Lazareff} visibility models, using a point source, an elliptical Gaussian in place of the disc and a ``halo''. This halo represents all the incoherent emission from outside regions. They found the contribution of the ellipse to be \(27\,\%\), the star \(67\,\%\), and the halo \(6\,\%\).

\citet{bouvier2020investigating} also performed multicolour photometry from the Las Cumbres Observatory Global Telescope Network (LCOGT) and spectroscopy using the  Canada France Hawaii Telescope (CFHT). They found a rotation period of \(P = 2.96\,\)days, a stellar inclination of \(i = (30 \pm 5)\degree\), which is in agreement with the inclination of the ellipse modeling (supposedly) the inner rim seen by VLTI. From their H\(\alpha\), H\(\beta\), Pa\(\beta\), Br\(\gamma\) spectra, they conclude that the spectral line profiles are best explained as a combination of funnel flows along magnetic field lines and disc winds, similar to the ones described by radiative transfer models by \citet{kurosawa2011multidimensional}, where the ``funnel'' flows cross the line of sight. For some spectral lines, this behaviour is periodic (He~I), and for others, they report a more complicated chaotic variability (Br\(\gamma\)).  \citet{bouvier2020investigating} derived an accretion rate of \(\dot{M} = 6.5\times10^{-9} {\,M_\odot}\,{\rm yr^{-1}}\).

\citet{2018ApJ...863...44A} studied discs around T Tauri stars (the DARTTS-S survey),
using the VLT/SPHERE/IRDIS instrument, operating between 0.9--2.3 microns, in the differential polarimetric imaging mode, with a coronagraph obscuring $0.1''$ around the central star. These images can be seen in Figures \ref{fig:multi}.A and \ref{fig:multi}.B, showing the scattered light from the surface of the disc. Surprisingly, they found no correlation of the scattered light, either to the stellar mass or to mm-fluxes. 

\citet{casassus2018inner} focused on the misalignment of the inner and outer discs of DoAr\,44. This is somewhat visible in the radio continuum, but it is very apparent in the polarimetric images. They estimated a tilt of \(30\degree\) of the inner disc with respect to the outer disc using radiative transfer modelling. 

\citet{casassus2019} construct a parametric radiative transfer model of DoAr\,44, accounting for the shadowing caused by the misaligned inner disc. They derive outer disc dust temperatures of 36 -- 48\,K with the cooler temperatures being in the nodal regions, shadowed by the inner disc. They derive a surface density of \(\Sigma = \SI{ <13}{\centi \meter^2 \gram^{-1}}\) and an aspect ratio of 0.08 at the inner rim of the outer ring.

\citet{arce2023radio} studied these same SPHERE images together with the ALMA Band~6 images from ODISEA and new Band~7 images, specifically of the DoAr\,44 system. They derived a relative inclination of \(21\degree\).

\citet{leiendecker2022dust} performed a study of DoAr\,44, together with the better known system HD 163296. They estimated a dust mass of 0.3\,\(M_{\rm Jupiter}\), corresponding to 21.5\,\(M_{\rm Jupiter}\) mass of gas. They tried explaining the ring morphology in alternative scenarios, such as a break-up of two large bodies. They also estimated the mass of a hypothetical planet needed to clear the inner disc at  \(0.5 \textrm{ to } 1.6\, M_{\rm Jupiter}\). 

In this work, we aim at constructing a global model of the system DoAr\,44 informed by a wide range of observables.

\section{Archival observations}

Several archival observations of DoAr\,44 are available, including
H\(\alpha\) spectra, VLTI interferometry, ALMA visibilities, and spectral energy distribution measurements.

\subsection{H$\alpha$ spectroscopy}\label{Halpha}

To constrain the physical properties of the accretion region around the star, all available H\(\alpha\) datasets from the ESO Archive were utilized. This includes 11 spectra from the VLT/ESPRESSO (program ID 111.250K), 3 spectra from VLT/UVES (program ID 69.C-0481) and a single spectrum taken by VLT/X-shooter (program ID 085.C-0764).

The VLT/ESPRESSO observations have a 2600\,s exposition time and have a spectral resolution of \(R = 140\,000\). The signal to noise ratio (SNR) in the continuum around H\(\alpha\) is better than \({\rm SNR} \gtrsim 50\), and reach \({\rm SNR} \approx 200\) in the center of the spectral line profile. Further details can be found in Tab.~\ref{tab:halfa}.

\begin{table}[]
\centering
\caption{Parameters of the VLT observations of the H\(\alpha\) spectrum of DoAr\,44. }\label{tab:halfa}
\begin{tabular}{lrrrl}
\hline \hline
Instrument & \multicolumn{1}{l}{\(R\)} & \multicolumn{1}{l}{SNR} & \multicolumn{1}{l}{Date} & \(t ~(s)\) \\ \hline
UVES & 66320 & 87.1 & 2020-10-20 & 900  \\
UVES & 66320 & 77.7 & 2020-10-20 & 1260  \\
UVES & 51690 & 55.2 & 2020-10-20 & 1800  \\
ESPRESSO & 140000 & 69 & 2023-05-05 & 2600  \\
ESPRESSO & 140000 & 72.9 & 2023-05-05 & 2600  \\
ESPRESSO & 140000 & 61.4 & 2023-05-05 & 2600  \\
ESPRESSO & 140000 & 42.8 & 2023-05-30 & 2600  \\
ESPRESSO & 140000 & 66.5 & 2023-05-05 & 2600  \\
ESPRESSO & 140000 & 81.1 & 2023-05-05 & 2600  \\
ESPRESSO & 140000 & 70 & 2023-05-05 & 2600 \\
ESPRESSO & 140000 & 73.5 & 2023-05-05 & 2600  \\
ESPRESSO & 140000 & 39.8 & 2023-05-30 & 2600  \\
ESPRESSO & 140000 & 64.1 & 2023-05-30 & 2600 \\
ESPRESSO & 140000 & 63.2 & 2023-05-05 & 2600 \\
X-shooter & 18340 & 300.3 & 2014-05-15 & 1120  \\ \hline
\end{tabular}
\tablefoot{$R$ denotes the spectral resolution,
SNR the signal to noise ratio,
and $t$ the exposure time.}
\end{table}

The H\(\alpha\) emission is extremely variable, changing from 6 to 12 times above the continuum emission, displaying single-, double-, triple-, and even quadruple-peaked profiles. This behaviour is extremely complicated, and is related to the detailed geometry and kinematics of the accretion funnel flows and accretion shocks, and the superposition of emission and absorption components \citep{bouvier2020investigating}.

\begin{figure}
\begin{minipage}[c]{0.45\textwidth}
    \includegraphics[width=\textwidth]{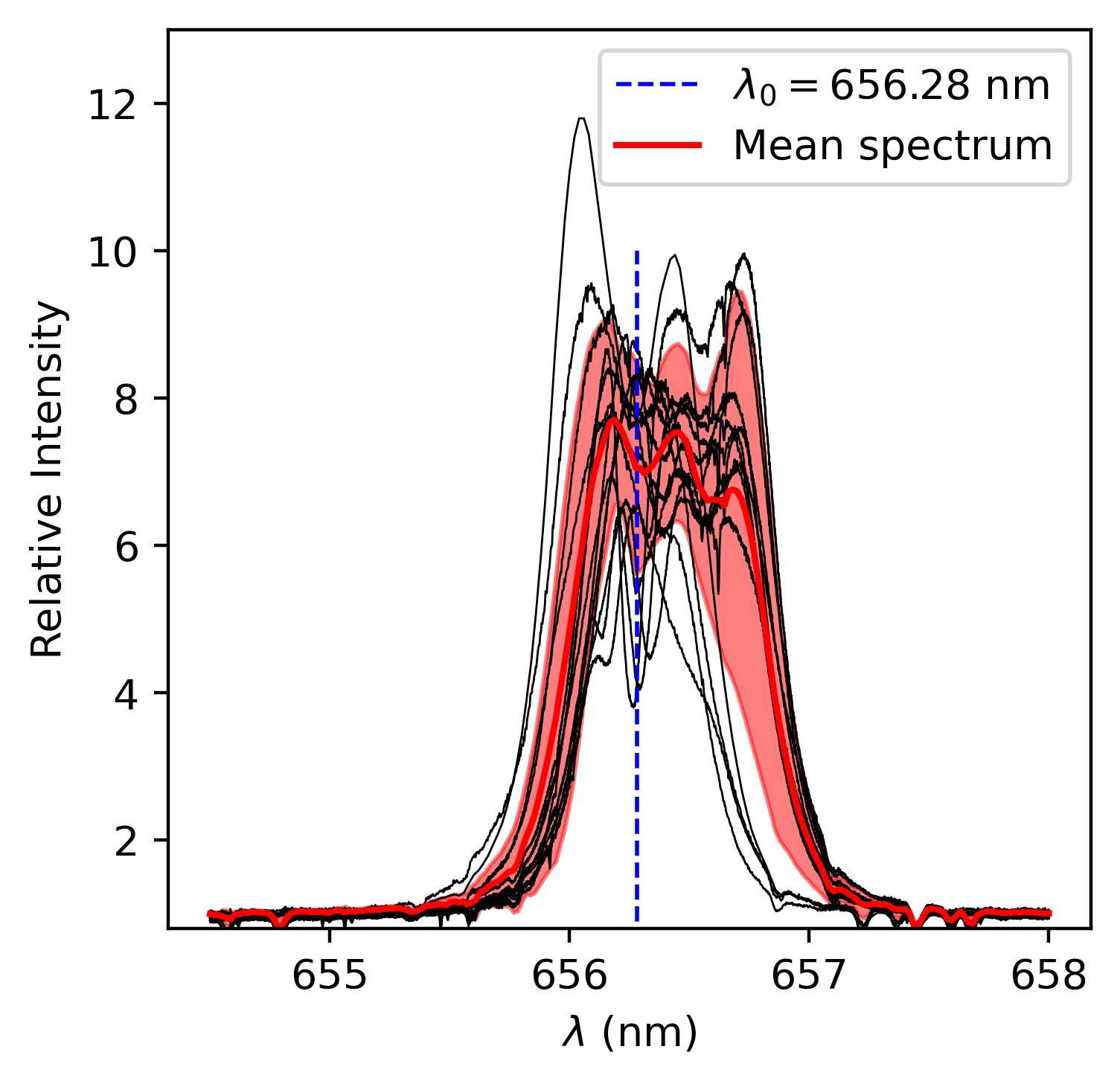}
\end{minipage}\hfill
\begin{minipage}[c]{0.45\textwidth}
    \caption{Individual VLT H\(\alpha\) spectra of DoAr\,44 and a mean spectrum. The spectrum is temporally variable, so a mean spectrum is used to constrain the mean state of the system.}  \label{fig:comb}
\end{minipage}
\end{figure}

For this reason, a mean profile was computed (see Fig. \ref{fig:comb}), corresponding approximately to an axially-symmetric mean state of the system.

A piece-wise linear interpolation has been performed in 200 equidistant points in the range 654.5 to \SI{658.0}{\nano \meter} for all the datasets, and a point-wise average and standard deviation from this average has been computed.  This standard deviation from the mean spectrum has been treated as the denominator in the \(\chi^2\) minimization, as our model will not capture the temporal variability of the system.

\subsection{VLTI interferometry}
\begin{figure}
\begin{minipage}[c]{0.24\textwidth}
    \includegraphics[width=\textwidth]{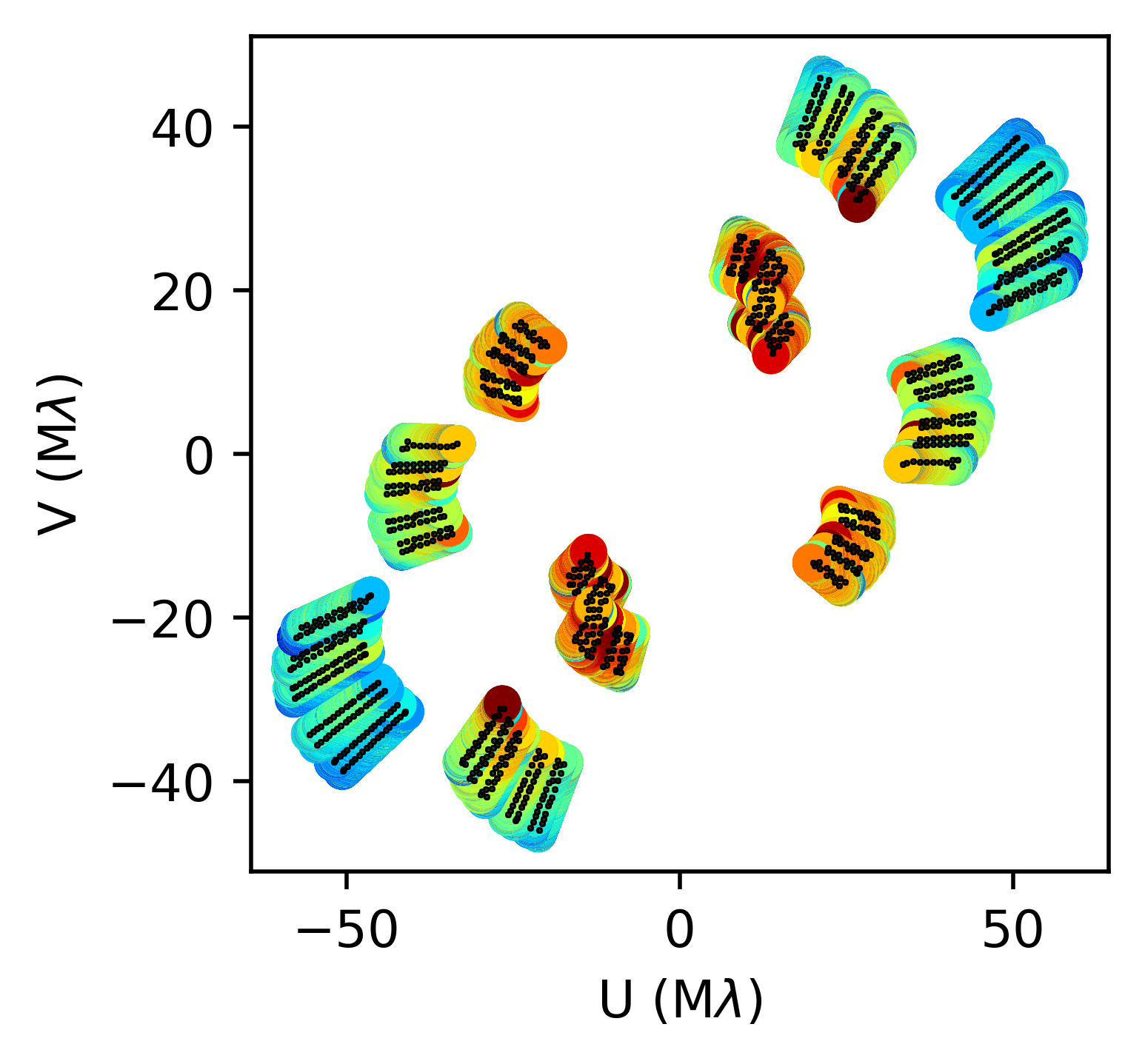}
    \end{minipage}\hfill
\begin{minipage}[c]{0.24\textwidth}
    \includegraphics[width=\textwidth]{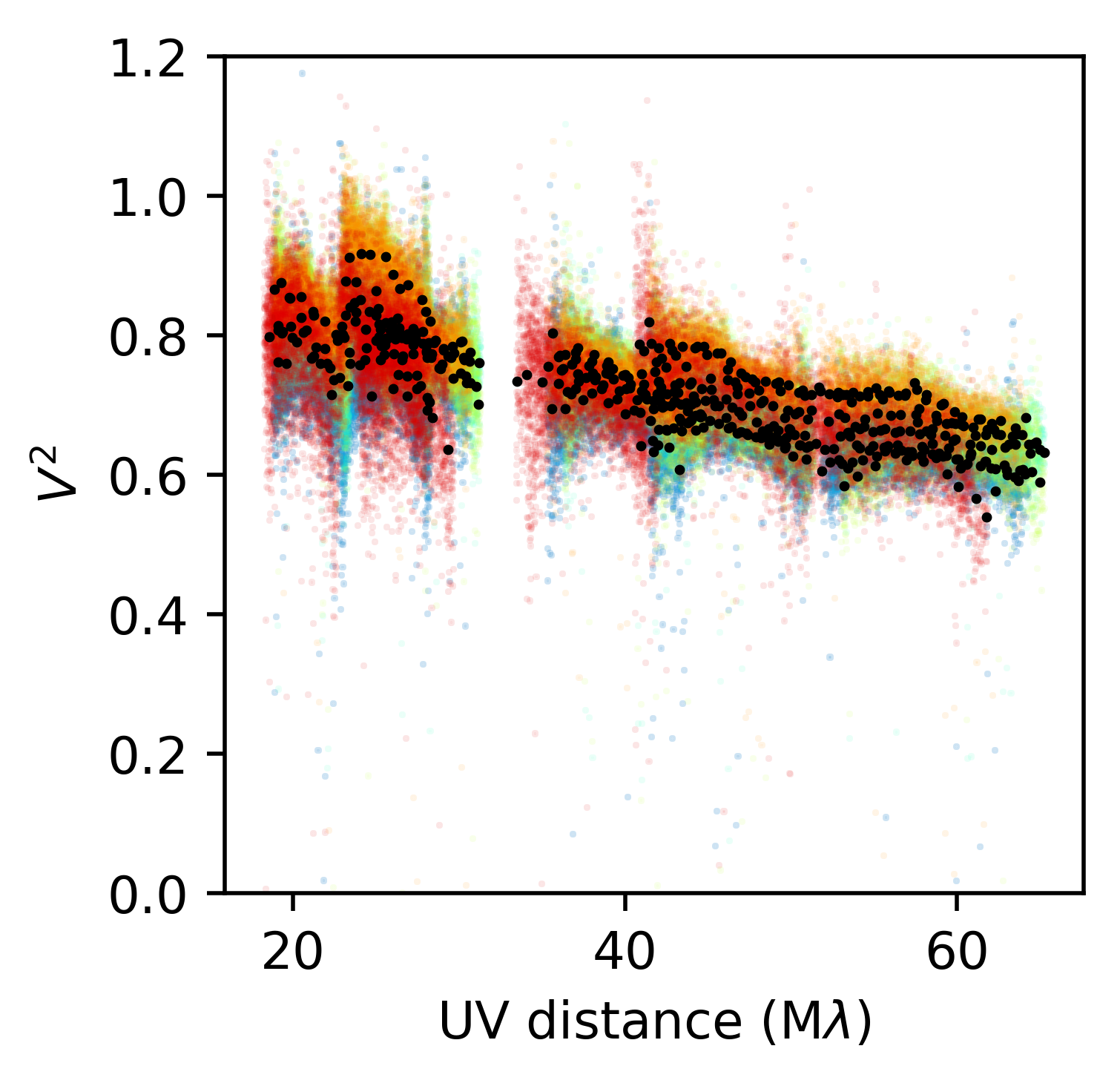}
\end{minipage}\hfill
\begin{minipage}[c]{0.45\textwidth}
    \caption[VLTI visibilites used for modelling.]{VLTI/GRAVITY K-band interferometric visibilities of DoAr\,44. \textbf{Left}: UV-coverage of the observations in M\(\lambda\). \textbf{Right:} The squared visibility as a function of UV distance (\(\sqrt{u^2 + v^2}\)) in M\(\lambda\). Notice that the \(V^2\) does not reach unity at the shortest baselines, indicating the presence of an extended component, whose angular size is larger than the maximum recoverable scale.}  \label{fig:VLTI}
\end{minipage}
\end{figure}

To constrain the spatial extent of the inner accretion region, we utilized archival VLTI/GRAVITY interferometric data of DoAr~44 \citep{bouvier2020probing}. The reduction was performed using the Gravity Pipeline in EsoReflex \citep{freudling2013automated}, to obtain the calibrated squared visibilities \(V^2\) and closure phases arg\(T^3\). The data was taken on the 22nd of June 2020 using the Unit Telescopes in the K-band (\(\approx \SI{2}{\micro \meter}\)). The uv-space coverage, together with the squared visibilities are shown in Fig. \ref{fig:VLTI}.

\subsection{ALMA interferometry}

\begin{figure}
\begin{minipage}[c]{0.45\textwidth}
    \includegraphics[width=\textwidth]{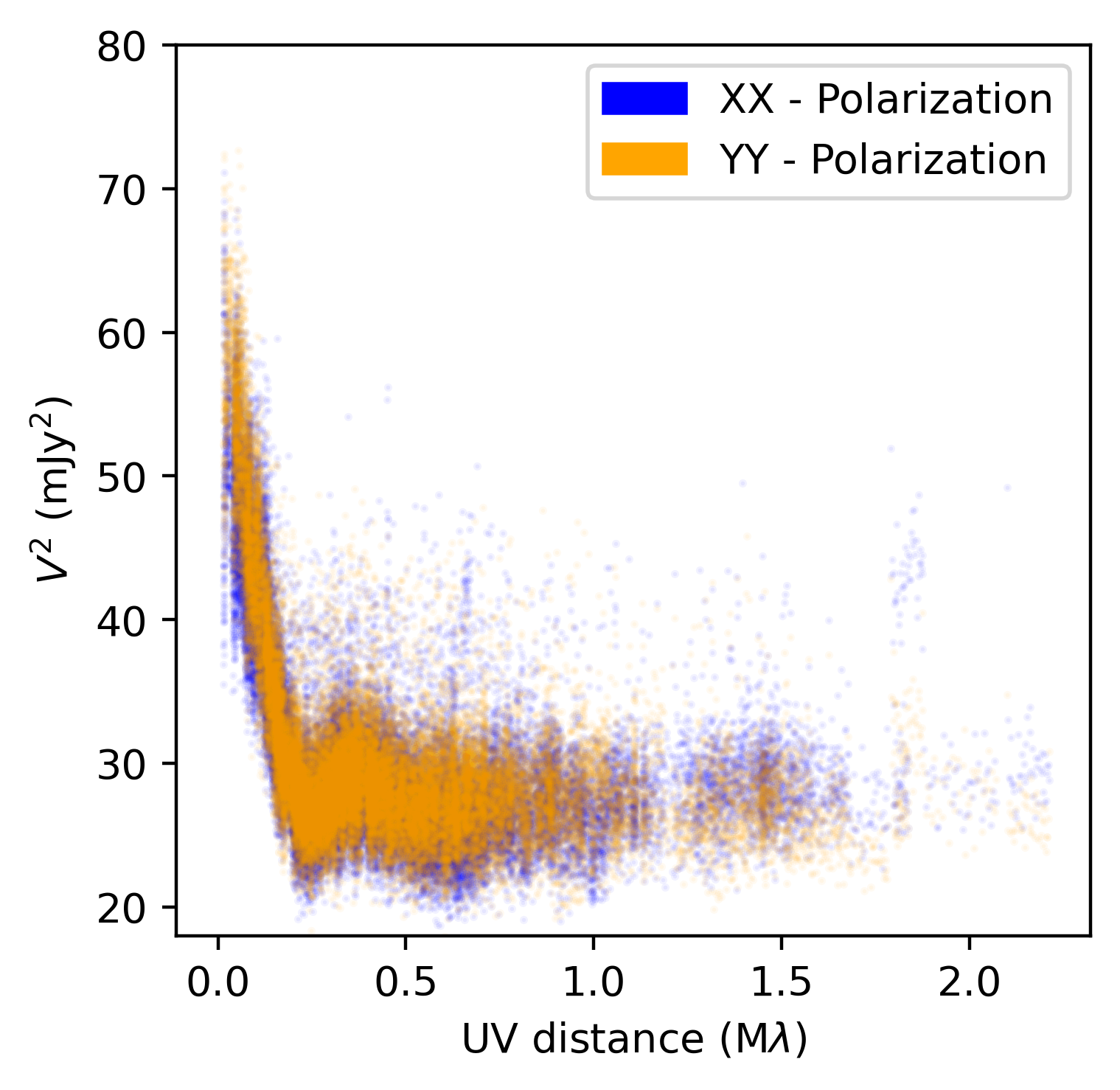}
\end{minipage}\hfill
\begin{minipage}[c]{0.45\textwidth}
    \caption[ALMA visibilites used for modeling.]{ALMA Band 7 Squared visibilites of DoAr\,44 for both polarizations (XX, YY) plotted separately as a function of uv-distance in M\(\lambda\).}  \label{fig:XXYY}
\end{minipage}
\end{figure}

To constrain the dimensions and temperature of the outer disc, the newest continuum Band 7 (\SI{870}{\micro \meter}) ALMA interferometric data was used (project code 2019.1.00532.S, observation date 2021-08-19, \citet{arce2023radio}). The minimum and maximum recoverable scales of the observations are  0.023" and 0.56" respectively, with a total 4390 s of integration time. The squared visibilities for both polarizations are shown in Fig. \ref{fig:XXYY}. The CLEANed image can be seen in \ref{fig:multi}.H.

A second high-resolution observation in ALMA Band 6 (project code 2018.1.00028.S, observation date 2020-11-30, \citet{cieza2021ophiuchus}) was also taken in consideration when modelling the intermediary disc. This observation can be seen in \ref{fig:multi}.G.

Other ALMA observations, listed in Table \ref{tab:ALMA_observations} have been used as total flux measurements, that is, as data points in the SED.

\begin{table*}
\centering
\caption[Available ALMA observations of DoAr\,44]{Currently available ALMA observations of DoAr\,44.}
\label{tab:ALMA_observations}
\begin{tabular}{lrllllll}
\hline \hline
Project code & \multicolumn{1}{l}{Band} & Frequency (GHz)  & Release date & minRS & maxRS  & t (s) \\ \hline
2019.1.01111.S & 3 & 89.533 -- 105.476 & 2022-08-13 & 0.251" & 3.80"  & 490  \\
2021.1.00378.S & 4 & 137.034 -- 152.977 & 2023-08-16 & 0.210" & 3.94" & 73   \\
2018.1.00028.S & 6 & 216.008 -- 233.992 & 2020-11-30 & 0.026" & 0.64"& 2830 \\
2012.1.00158.S & 7 & 329.093 -- 343.083  & 2015-08-19 & 0.229" & 1.98"& 2056  \\
 2019.1.00532.S & 7 & 342.417 -- 356.969  & 2022-06-08 & 0.168"& 2.70" & 2570  \\
2019.1.00532.S & 7 & 342.423 -- 356.969 & 2023-09-06 & 0.023" & 0.56" & 4391 \\
2021.1.00378.S & 8 & 397.050 -- 412.995 & 2023-09-01 & 0.154"  & 2.36" & 113   \\ \hline
\end{tabular}
\tablefoot{
The value of minRS corresponds to the minimum recoverable scale, i.e., the angular resolution, and the value maxRS denotes the maximum recoverable scale. The integration time is seconds is denoted by \(t\).
}
\end{table*}

\subsection{Spectral energy distribution}
The measurements of the SED were taken from the following surveys, publications and catalogues,
using mostly Vizier \citep{ochsenbein2000vizier}:

{\footnotesize

\begin{itemize}
\setlength\itemsep{0em}

    \item XMM-Newton, \(\SI{344}{\nano \meter}\) \citep{page2012xmm};
    
    \item ASAS-SN catalogue, \(\SI{444}{\nano \meter}\) -- \(\SI{2.16}{\micro \meter}\) \citep{jayasinghe2018asas};
    
    \item SkyMapper catalogue, \(\SI{497}{\nano \meter}\), \(\SI{604}{\nano \meter}\) \citep{SkyMapper};
    
    \item Gaia satellite, \(\SI{505}{\nano \meter}\), \(\SI{673}{\nano \meter}\) \citep{gaia2018vizier, gaia2020};
    
    \item AAVSO Photometric All Sky Survey (APASS), \(\SI{482}{\nano \meter}\) -- \(\SI{763}{\nano \meter}\) \citep{2016yCat.2336....0H};
    
    \item Sloan Digital Sky Survey (SDSS), \(\SI{482}{\nano \meter}\), \(\SI{763}{\nano \meter}\) \citep{york2000sloan};
    
    \item Pan-STARRS survey, \(\SI{613}{\nano \meter}\) -- \(\SI{960}{\nano \meter}\) \citep{kaiser2002pan};
    
    \item 2MASS, \(\SI{1.65}{\micro \meter}\), \(\SI{2.16}{\micro \meter}\) \citep{skrutskie2006two};
    
    \item Wide-field Infrared Survey Explorer (WISE), \(\SI{3.35}{\micro \meter}\), \(\SI{11.6}{\micro \meter}\) \citep{wright2010wide};
    
    \item Spitzer/IRAC, \(\SI{3.6}{\micro \meter}\) -- \(\SI{7.8}{\micro \meter}\) \citep{fazio2004infrared};
    
    \item Spitzer/MIPS, \(\SI{70}{\micro \meter}\) \citep{rieke2004multiband};
    
    \item IRAS satellite, \(\SI{11.6}{\micro \meter}\), \(\SI{61}{\micro \meter}\) \citep{neugebauer1984infrared};
    
    \item AKARI satellite, \(\SI{18}{\micro \meter}\), \(\SI{90}{\micro \meter}\) \citep{2010A&A...519A..83T};
    
    \item Herschel/PACS, \(\SI{100}{\micro \meter}\), \(\SI{160}{\micro \meter}\) \citep{billot2006herschel};
    
    \item Herschel/SPIRE, \(\SI{250}{\micro \meter}\), \(\SI{350}{\micro \meter}\) \citep{griffin2010herschel};
    
    \item JCMT, \(\SI{850}{\micro \meter}\) \citep{2013ApJ...773..168M};
    
    \item ATCA, \(\SI{3}{\milli\meter}\) \citep{ricci2010dust}.

\end{itemize}

}

Moreover the integrated ALMA photometry was used, together with the interferometric observations described below. The X-ray data from \citet{montmerle1983einstein} were not used, as the X-ray emission is formed under non-LTE conditions, which is outside the approximations of our models.

\section{Simple analytic models}\label{sec:analmod}

Before we begin constructing radiative-transfer models, it will be useful to construct a few simple analytic models to act as benchmarks for our later work. With these analytic models, we aim to approximately describe the observed interferometric visibilities. These models are as follows:
\begin{enumerate}
    \item A homogeneous ellipse described by \(\alpha, \beta, \omega\), together with a point source and a background halo for the VLTI/GRAVITY visibilities, to model the inner disc.
    \item A narrow elliptical ring with a background halo for the ALMA visibilities, to model the outer disc.
\end{enumerate}

Similar models were used by \citet{Lazareff} to model VLTI/PIONIER visibilities of about 50 Herbig AeBe discs. The introduction of a halo allows us to lower the visibility at short baselines that are not sampled by the VLTI. However, this contribution is not merely a mathematical aid, but is directly observed by coronagraphic instruments, such as the VLT/SPHERE -- see Figure \ref{fig:multi}. The two models can be described as follows:
\begin{align}\label{eq:analmod}
    V_1(u, v) &= \frac{f_{\rm s} + f_{\rm c} \mu_{\rm e}(u, v;  \alpha, \beta, \omega)}{f_{\rm h} + f_{\rm e} + f_{\rm s}}, \\
    V_2(u, v) &= \sqrt{A + B\mu_\delta^2(u, v; \alpha, \beta, \omega)}\,,
\end{align}
where the visibility of an ellipse, describing the circumstellar disc, is:
\begin{equation}
    \mu_{\rm c} (u, v; \alpha, \beta, \omega) = \frac{2 J_1\left(2\pi \sqrt{\alpha^2 u'^2 + \beta^2 v'^2} \right)}{2\pi \sqrt{\alpha^2 u'^2 + \beta^2 v'^2}}\,,
\end{equation}
and of a \(\delta\)-ring is:
\begin{equation}
        \mu_{\rm \delta}(u, v; \alpha, \beta, \omega) = J_0\left(2\pi \sqrt{u'^2 \alpha^2 + v'^2 \beta^2}\right)\,,
\end{equation}
with the rotated aperture-plane coordinates given by:
\begin{equation}
\begin{pmatrix}
u' \\
v'
\end{pmatrix}
=
\begin{pmatrix}
\cos\omega & +\sin\omega \\
-\sin\omega & \cos\omega
\end{pmatrix}
\begin{pmatrix}
u \\
v
\end{pmatrix}\,.
\end{equation}
The parameters \(f_{\rm c}, f_{\rm h}, f_{\rm s}\) denote the fractional contributions of the circumstellar disc, the halo and the star. The coordinates \(u, v\) and the angular sizes \(\alpha\) and \(\beta\) are dimensionless, that is, given in cycles per baseline and in radians, respectively. As the separations of the telescopes (or antennas) is much larger than the wavelength that we are observing, we commonly use dimensionless units such as M\(\lambda\) to denote \(10^6\) cycles per baseline. Best-fit parameters are estimated in Tables \ref{tab:modelB} and \ref{tab:modelC}.

\begin{table}[ht]
\centering
\caption[Parameters of best-fit MCMC star and ellipse and halo model suitable for inner accretion region for VLTI visibilities]{Best-fit parameters of the homogeneous ellipse model suitable for the VLTI measurements of the accretion region. }
\vspace{1em}
\label{tab:modelB}
\begin{tabular}{ccc}
\hline \hline
Parameter & value & \(\sigma\) \\ \hline
\(\alpha\) (mas) & 1.243 & 0.2 \\
\(\beta\) (mas) & 1.663 & 0.3 \\
\(\omega\) \((^\circ)\) & 147 & 2 \\
\( f_{\rm s}\)& 0.68 & 0.1 \\
\( f_{\rm c}\)& 0.24 & 0.1 \\
\( f_{\rm h}\)& 0.072 & 0.005 \\
\hline
\end{tabular}
\tablefoot{The linear scale of the accretion region is 0.24\,au.}
\end{table}

\begin{table}[ht]
\centering
\caption[Parameters of best-fit MCMC \(\delta\)-ring model of the outer disc for ALMA visibilities.]{Best-fit parameters of the elliptical \(\delta\)-ring model, suitable for the ALMA visibilities.  }
\vspace{1em}
\label{tab:modelC}
\begin{tabular}{ccc}
\hline \hline
Parameter & value & \(\sigma\) \\ \hline
\(\alpha\) (mas) & 329 & 1 \\
\(\beta\) (mas) & 304 & 1 \\
\(\omega\) \((^\circ)\) & 156 & 2 \\
\( A ~( 10^5 \rm{Jy}^2)\) & 3.721 & 0.003 \\
\( B ~( 10^5 \rm{Jy}^2)\) & 3.827 & 0.012 \\
\hline
\end{tabular}
\tablefoot{The \(A\) and \(B\) parameters were used to rescale the visibilities in physical units (Jy) into a unit-less quantity for further analysis.}
\end{table}

These analytic models were used as a benchmark for the more complex radiative-transfer models, providing the angular extent of the observed structures as a guide. Also, they provided a way to deal with the fully resolved component in the VLTI observations, and with the fact that the ALMA observations are in physical and not in relative units. The interferometric quantities were rescaled in the following way:
\begin{align}
    V^2_{ \rm VLTI,new} &=\frac{1}{e^2} V^2_{\rm VLTI}\,,\\
    T_{3\,,\rm VLTI,new} &=\frac{1}{e^3} T_{3\,,\rm VLTI}\,,\\
    V^2_{\rm ALMA,new} &=-A + \frac{1}{B} V^2_{\rm ALMA}\,,
\end{align}
where \(e = 0.928\) from the homogeneous ellipse model, and \(A, B\) are taken from the elliptical \(\delta\)-ring model. The best-fit parameters and their uncertainties (68\% confidence intervals of the marginalized posterior distributions) were obtained via a MCMC simulation with an uninformative prior. In each case, two solutions were obtained for two position angles separated by $180^\circ$, and only the solution with \(\omega < 180^\circ\) was considered. Initial simulations were performed with 32 walkers and \(10^4\) steps. The marginalized posteriors were used as a starting value for the next iteration: these new simulations ran with 64 walkers for a total of \(10^5\) steps, with a \( 30\%\) burn-in phase. All simulations were implemented using the \texttt{emcee} package \citep{foreman2013emcee}.

From all these models, we infer an inclination of the inner accretion region at
\(i_{\rm in} = 40\degree \pm 3 \degree\)
and of the outer disc at
\(i_{\rm out} = 22\degree \pm 1\degree\).
The positions angles seem to be somewhat aligned, at
\(\omega_{\rm in} = 147\degree\pm 2\degree\)
for the inner accretion region and
\(\omega_{\rm out} = 156\degree \pm 2\degree\)
for the outer disc. 

The misalignment \(\xi\) can be calculated, for example, using the spherical law of cosines:
\[ \cos \xi = \cos i_{\rm in} \cos i_{\rm out}  + \sin i_{\rm in} \sin i_{\rm out} \cos(\Delta \omega) = 18.5 \degree \pm 4 \degree\,.\]

For comparison, \citet{bohn2022probing} investigated the inner and outer disc misalignment of a series of protoplanetary discs, including DoAr\,44, also based on VLTI/GRAVITY and ALMA interferometry. For DoAr\,44, they derived an inner disc inclination of 
\(i_{\rm in} = \left(25.67^{+7.91}_{-9.71}\right) \degree\)
and an outer disc inclination of
\( i_{\rm out} = \left(23.20^{+1.98}_{-1.58}\right) \degree\). 

\citet{casassus2018inner} used VLT/SPHERE/IRDIS polarimetric measurements and ALMA Band 7 continuum measurements to derive a misalignment of
\(\xi = 30\degree \pm 5 \degree\).

\citet{bouvier2020probing} derived an inner disc inclination of
\(i_{\rm in} = 34\degree \pm 2 \degree\)
based on VLTI/GRAVITY interferometric visibilities, with a position angle of
\(\omega_{\rm in} = 140\degree \pm 3\degree \).

\citet{arce2023radio} derived outer disc inclinations for two high-resolution ALMA Band 6 and Band 7 observations,
\(i_{\rm out} = 21.8\degree \pm0.3\degree\).
Alternatively, they derive an outer disc inclination of
\(24.4\degree \pm0.4\degree\)
using H-band polarised intensity images from VLT/SPHERE/IRDIS.
They derive an inner-outer misalignment of
\(\xi = \left(21.4^{+6.7}_{-8.3}\right)\degree\).

In conclusion, our results are in agreement for the outer inclination (1 sigma),
as well as for the new value of misalignment.
The inner inclination is more uncertain, as the differences between models seem to be of the order of 2 sigma.

\section{\texttt{Pyshellspec} radiative-transfer models}

The primary tool used to create radiative transfer models for this work,
\texttt{Pyshellspec}, was developed by \citet{pyshellspec2018} and \citet{pyshellspec}
and to model the $\beta$ Lyr\ae~A system.
\texttt{Pyshellspec} generates various observables, such as interferometric visibilities and closure phases, based on synthetic images generated by \texttt{Shellspec}, a long-characteristics radiative-transfer code by \citet{shellspec2004}. \texttt{Shellspec} is used for modelling binary stars moving within circumstellar matter (i.e., gas and dust), and includes a series of parametrically defined objects, such as stars, rings, discs, jets or spherical shells, which shall be later described. 

The radiative-transfer equation is integrated along the line of sight:
\begin{equation}
    {\rm d} I_\nu = (\epsilon_\nu - \chi_\nu I_\nu) {\rm d}x\,,
\end{equation}
where \(\epsilon_\nu, \chi_\nu\) are the total emissivity and the total opacity of the medium, \(I_\nu\) is the specific monochromatic intensity, and \({\rm d} x\) is measured along the ray.

The total opacity is given as the sum of the absorption opacity \(\kappa_\nu\) and the scattering opacity \(\sigma_\nu\):
\begin{equation}
     \chi_\nu = \kappa_\nu + \sigma_\nu\,.
\end{equation}
\texttt{Shellspec} allows the user to include into \(\kappa_\nu\) the following sources of absorption opacity:
\begin{itemize}
    \setlength\itemsep{0em}
    \item Spectral line opacity,
    \item HI (neutral hydrogen) bound-free opacity,
    \item HI free-free opacity,
    \item H\(^-\) (negative hydrogen ion) bound-free opacity,
    \item H\(^-\) free-free opacity,
    \item Mie absorption on dust,
\end{itemize}
and the following sources of scattering opacity:
\begin{itemize}
    \setlength\itemsep{0em}
    \item Thompson (i.e., low-energy Compton) scattering on free electrons,
    \item Rayleigh scattering on neutral hydrogen,
    \item Mie scattering on dust.
\end{itemize}
The dust opacities are summed over all the included dust species that are located in a cold enough region, allowing for dust condensation.

The total emissivity is calculated as the sum of the thermal emissivity and the emission due to scattering:
\begin{equation}
    \epsilon_\nu = \epsilon^{\rm th}_\nu + \epsilon^{\rm sc}_\nu = B_\nu(T) \kappa_\nu + \epsilon^{\rm sc}_\nu\,,
\end{equation}
where \(B_\nu\) is the Planck function.
One limitation is that \texttt{Pyshellspec} is a single-scattering radiative-transfer tool.
That means that the stellar light scattered off the disc is accounted for in the synthetic image. 
We tested models with and without scattering opacity. The absorptive
term leads a decrease of the SED at mid-infrared wavelengths, however,
this should be at least partly compensated by the emissive term,
scattering thermal radiation from other directions, if it is accounted for
and if the intensity is approximately isotropic. According to these tests,
we consider our nominal models as good approximations.

For the evaluation of the goodness of fit, a standard \(\chi^2\) metric is defined:
\begin{equation}
    \chi^2 = \chi^2_{\rm SPE} + \chi^2_{\rm V2} + \chi^2_{\rm CP} + \chi^2_{\rm T3} + \chi^2_{\rm SED}\,,
\end{equation}
\noindent
where
\(\chi^2_{\rm SPE}\) is the \(\chi^2\) contribution from the spectroscopic data,
\(\chi^2_{\rm V2}\) the squared visibilities,
\(\chi^2_{\rm CP}\) the closure phases,
\(\chi^2_{\rm T3}\) the triple product amplitudes, and
\(\chi^2_{\rm SED}\) the SED. These are defined as:
\begin{align}
    \chi^2_{\rm SPE} &= \sum_{i = 1}^{N_{\rm SPE}} \left(\frac{I^i_{\lambda\,, \rm obs} - I^i_{\lambda\,, \rm syn}}{\sigma_i}\right)^2\,, \\
    \chi^2_{\rm V2} &= \sum_{i = 1}^{N_{\rm V2}} \left(\frac{|V^i_{\rm obs}|^2 - |V^i_{\rm syn}|^2}{\sigma_i}\right)^2 \\
    \chi^2_{\rm CP} &= \sum_{i = 1}^{N_{\rm CP}} \left(\frac{ \mathrm{arg}\,(T_3^i)_{\rm obs} -\mathrm{arg}\,(T_3^i)_{\rm syn} }{\sigma_i}\right)^2\,,\\
    \chi^2_{\rm T3} &= \sum_{i = 1}^{N_{\rm T3}} \left(\frac{ |T_3|^i_{\rm obs} -|T_3|^i_{\rm syn} }{\sigma_i}\right)^2\,,\\
    \chi^2_{\rm SED} &= \sum_{i = 1}^{N_{\rm SED}} \left(\frac{ F^i_{\lambda \,, \rm obs} - F^i_{ \lambda \,, \rm syn} }{\sigma_i}\right)^2\,,
\end{align}
where
$I_{\lambda\,,{\rm obs}}$ and $I_{\lambda\,,{\rm syn}}$ denote the observed and synthetic monochromatic intensities, respectively,
$V$ the visibilities,
${\rm arg}\,T_3$ the closure phases,
$|T_3|$ the triple product amplitudes, and
$F_{\lambda}$ the monochromatic fluxes.

However, the total \(\chi^2\) was not optimised globally, as this is exceedingly difficult.
The only contribution optimised globally was \(\chi^2_{\rm SED}\),
while the other contributions were optimised separately for the accretion region, the inner disc, and the outer disc.

Our model comprises three main parts:
\begin{enumerate}
    \item the accretion region, described in our model using a star, a shell of infalling gas and a hot innermost disc (up to 0.1\,au),
    \item an intermediary disc (0.2 - 1\,au), and
    \item the outer disc (40 - 60\,au).
\end{enumerate}

The discs and the spherical shell were modelled using \texttt{Shellspec}'s \texttt{DISC} and \texttt{SHELL} objects, respectively.

\subsection{The central star}

The simplest element in our model is the central star. A synthetic stellar spectrum is provided as a boundary condition. This stellar spectrum was obtained by interpolation in the PHOENIX grid \citep{husser2013new}, similarly to that in \citet{nemravova2016xitauri}. The interpolation was performed in the 3-dimensional parameter space of the effective temperature \(T_{\rm eff}\), the (logarithm of) surface gravity \(\log g \) and the metallicity \(Z\). The stellar parameters were set according to previously published works \citep{andrews2009protoplanetary, bouvier2020probing},

$r_\star = 2.0\,R_\odot$,
$T_\star = 4750\,{\rm K}$,
$M_\star = 1.4\,M_\odot$.

\subsection{The \texttt{DISC} object}

The primary object used to model discs in \texttt{shellspec} is the \texttt{DISC}. The disc is parameterized by the following equations:  

\begin{align}
\Sigma(R) &= \Sigma(R_{\rm innb}) \left(\frac{R}{R_{\rm innb}}\right)^{e_{\rm densnb}}, \\
H(R) &= R \frac{c_s(R)}{v(R)}, \\
c_s(R) &= \sqrt{\frac{\gamma k_{\rm B} T(R)}{\mu}}, \\
T(R) &= T_{\rm nb} \left(\frac{R}{R_{\rm innb}}\right)^{e_{\rm tempnb}}, \\
\varrho_{\rm dstnb} &= Z \varrho_{\rm nb}.
\end{align}
where:  
\begin{itemize}
    \item \(\Sigma(R_{\rm innb})\), \(R_{\rm innb}\), and \(e_{\rm densnb}\) describe the surface density, inner radius, and density power-law exponent, respectively.
    \item \(H(R)\) represents the disc's scale height.
    \item \(v(R)\) denotes the Keplerian velocity.
    \item \(c_s\) is the speed of sound, with \(\gamma = c_p / c_v\) (the ratio of specific heats) and \(\mu\) (the mean molecular weight).
    \item \(T(R)\) describes the temperature profile, parametrized by \(T_{\rm nb}\) (the temperature normalization) and \(e_{\rm tempnb}\) (the temperature exponent).
    \item \(\varrho_{\rm dstnb}\) represents the dust density in the protoplanetary disc, linked to the gas density \(\varrho_{\rm nb}\) through a constant (solar) metallicity \(Z = 0.0122\) \citep{ASPLUND20061}.
\end{itemize}

The model accounts for scattering and absorption opacities due to dust particles of various sizes and mineral compositions. A mix of 90\% water ice and 10\% forsterite is used as a proxy for the actual mineralogical composition. This is a slightly higher ratio than the traditional 80 \% ices to 20 \% refractory materials \citep{Dodson-Robinson_2011, pontoppidan2014volatiles}. Detailed dust evolution modelling suggests that virtually all dust grains in the outer disc are covered in ice \citep{krijt2016tracing}, impacting its optical properties \citep{kohler2015dust, ysard2018optical}. We account for this fact using an increased water ice content.

\subsection{The \texttt{SHELL} object}
The velocity \(v(r)\), density and temperature profiles in the \texttt{SHELL} are described as follows:
\begin{equation}
    v(r)=v_0\left(1-\frac{r_{\rm csh}}{r}\right)^{e_{\rm velsh}}\,,
\end{equation}
\begin{equation}
    \rho(r)=\varrho_{\rm sh} \left(\frac{r_{\rm insh}}{r}\right)^2 \frac{v(r_{\rm insh})}{v(r)}\,,
\end{equation}
\begin{equation}
    T(r) =T_{\rm sh}=\rm{const.}\,,
\end{equation}
 where \(r_{\rm csh}, r_{\rm insh},e_{\rm denssh}\) are parameters. The shell is truncated at an outer radius \(r_{\rm outsh}\). To obtain a contribution to the opacity due to Thompson scattering, the electron number density is computed using the Saha equation based on the local densities and the temperature, with the assumption of LTE.

\subsection{Accretion region model}
\begin{figure*}
    \centering
  \includegraphics[width=\textwidth]{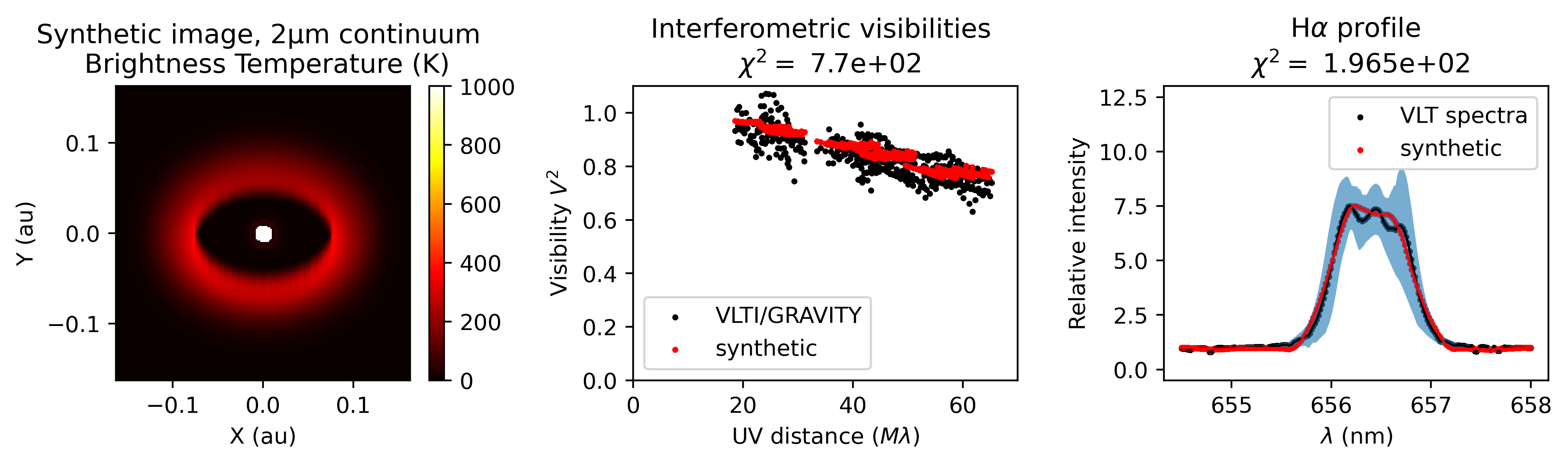}
    \caption[Best-fit models of inner accretion region]{\textbf{Left:} Synthetic image of the accretion region, in the \SI{2}{\micro\meter} continuum. \textbf{Center:} Fit of VLTI/GRAVITY squared visibilities. \textbf{Right:} Fit of the averaged H\(\alpha\) profile. The blue region is the standard deviation of the VLT spectra, representing the temporal variability.}
    \label{fig:inner_shell_disc_model}
\end{figure*}

The geometry that best replicates the observables is the following: a central star, surrounded by a spherical shell of hot gas, surrounded by a disc. The geometry of the disc was taken from the MCMC analysis of the analytic models, the inclination of \(i = 40\degree\) is fixed, together with the position angle of 147\(\degree\). The comparison of the spectroscopic and interferometric VLT+VLTI measurements with our synthetic data is shown in Figure \ref{fig:inner_shell_disc_model}.

A search for the best-fit was performed on a sub-space of the parameter space that was identified (via trial and error) as being spanned by the most relevant parameters:
\begin{enumerate}
    \item \(R_{\rm outsh}\), the outer radius,
    \item \(v(R_{\rm insh})\), the velocity at the inner boundary,
    \item \(e_{\rm velsh}\), the exponent of the power-law describing the velocity field of the shell,
    \item \(T_{\rm sh}\), the (constant) temperature,
    \item \(\varrho_{\rm sh}\), the density at the inner boundary.
\end{enumerate}
The \(R_{\rm insh}\) had to be fixed slightly above the stellar surface (for not to cause a collision with the \texttt{STAR} object).

The central star dominates the flux contribution in the optical domain. The star is surrounded by an envelope of optically thin, hot (\(\approx \SI{9000}{\kelvin}\)) gas, with a density of (\(\approx \SI{2e-13}{\gram \per \cubic \centi \meter}\)). This envelope extends up to about 0.1\,au, where it encounters the innermost disc, and is responsible for the majority of the H\(\alpha\) emission. The radial velocity of the gas in this envelope reaches \(v = \SI{385}{\kilo \meter \per \second}\) The inner rim of this disc is the main source of near-infra-red continuum radiation, and is responsible for the observed VLTI/Gravity visibilities. 

The model was fitted to the VLTI/Gravity observations, the VLT \(H\alpha\) spectra. The respective normalized \(\chi^2\) contributions are \(\chi^2_{\rm SPE} / N_{\rm SPE} = 0.98\) and \(\chi^2_{\rm V2} / N_{\rm V2} = 1.5 \).
The best-fit values of these parameters are listed in Tab. \ref{tab:bestfit}.

From the density and velocity of the gas in the accretion envelope shell, we derive an accretion rate of:
\begin{equation}
    \dot{M} = (6.1\pm 0.2)\cdot10^{-8}\,M_\odot\, {\rm yr}^{-1}\,,
\end{equation}
which is surprisingly similar to the result obtained by \citet{bouvier2020investigating}.

As the temperatures and spatial scales of this accretion region clearly do not account for the SED at mm-wavelengths and ALMA visibilities, we must model the cold outer disc.

\subsection{Outer disc model}

To account for the ALMA visibilities and CLEANed images, we were forced to introduce a cavity. A scan over a variety of parameters (particle size, temperature, density, inner radius, outer radius) fitting the \(\chi_{\rm V2}^2 + \chi^2_{\rm SED}\) was performed. The constraint on the total area of the disc is given primarily from the ALMA interferometry, where the inner and outer boundaries are visible (up to the background RMS level, the dust extends up to \(\approx 80\)\,au, see Fig. \ref{fig:multi}). The density profile was simplified to a power-law, and the complex ``ring-within-a-ring" morphology was not fitted in detail.

In our model, the outer disc extends from 40\,au to 60\,au, with the mid-plane dust density of $\SI{1e-14}{\gram \per \cubic \centi \meter}$ at the inner boundary. The dust is given by a mix of water ice and forsterite, in a ratio of 90 \% and 10 \%, with particle sizes of \(a = \SI{1}{\micro \meter}\). The mid-plane temperature at the 40\,au inner boundary is $\SI{60}{\kelvin}$, and decreases as a power-law with an exponent of $-0.5$.

\begin{figure*}
    \centering
  \includegraphics[width=\textwidth]{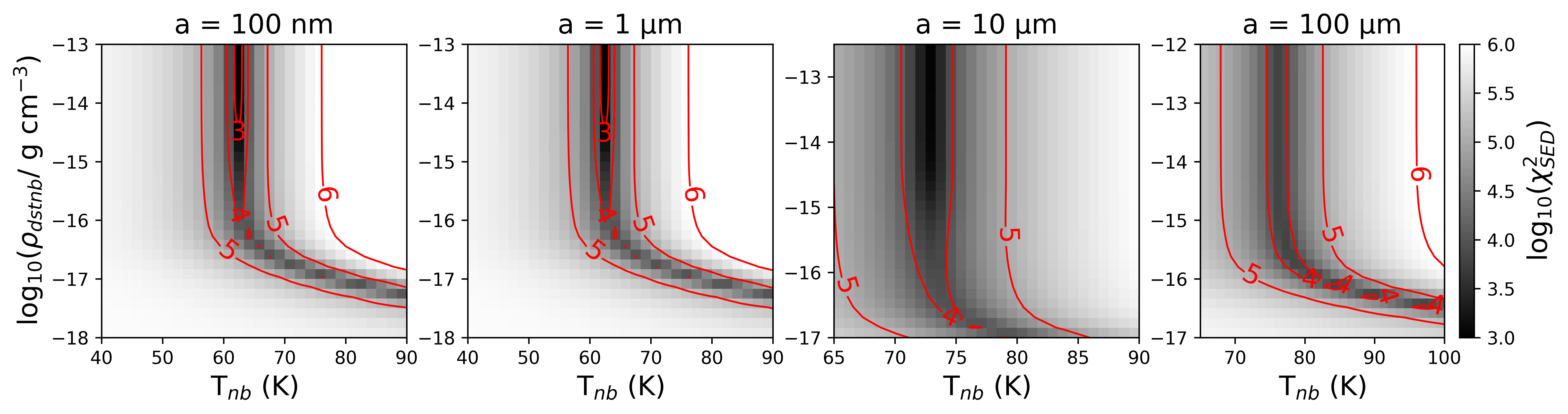}
    \caption[The \(\chi^2_{\rm SED}\) contribution of the outer disc of DoAr\,44 for a variety of dust sizes.]{The \(\chi^2_{\rm SED}\) contribution of the outer disc, for a variety of different particle sizes, from 100\,nm to \(\SI{100}{\micro \meter}\). The curved low-density high-temperature sections correspond to optically thin solutions, and the vertical (isothermal) dark areas correspond to optically thick solutions.}
    \label{fig:dust_comparisons_systematic}
\end{figure*}

The aspect ratio $H/r$ is 0.08 for the inner region of the outer disc (40\,au),
in agreement with \citet{casassus2018inner},
which then gradually increases to 0.10 in the outermost regions (60\,au).
By integrating the vertical density profile,
we obtain the dust surface density of the order of
$\Sigma_{\rm dust} \gtrsim 0.5\,{\rm g}\,{\rm cm}^{-2}$.
By integrating over the radial direction, we obtain the dust mass of
\begin{multline}
M_{\rm dust} = \int_{R_{\rm in nb}} ^{R_{\rm outnb}} 2 \pi r \Sigma_{\rm dust}(r) {\rm d} r= \\ = \int_{R_{\rm in nb}} ^{R_{\rm outnb}} 2 \pi r \varrho_{\rm dstnb} r \sqrt{\frac{r}{GM}} \sqrt{\frac{\gamma k_{\rm B} T(r)}{\mu}} {\rm d} r \gtrsim 100 M_{\rm Earth}\,.
\end{multline}
This is still a lower limit though, as the outer disc in our model is optically thick: the best-fit dust density estimations do not have a clear upper limit from fitting the SED, as various optically thick solutions appear very similar; see Fig.~\ref{fig:dust_comparisons_systematic}).

For comparison, \citet{cieza2021ophiuchus} assumed a single value of temperature of \(T = 20\,{\rm K}\) and derived \(M_{\rm dust} = 54\,M_{\rm Earth}\) for DoAr\,44, which is again a lower limit.
Our model requires an optically thick outer disc in the \(\SI{870}{\micro \meter}\) continuum, which is necessary to explain the SED and the sub-mm spectral index of $\alpha\approx 2$. The work \citet{arce2023radio} also suggests that the disc is optically thick, based on the well-constrained spectral index.

\citet{leiendecker2022dust} derived a dust mass of \(84^{+7.0}_{-3.5}\,M_{\rm Earth}\) for the outer disc, based on ALMA Band 6 observations. Their mid-plane temperatures are around \(30\,{\rm K}\), and they increase with increasing height above the mid-plane. On the contrary, our model is vertically isothermal, and requires about 50 -- 60\,K in the outer regions, in order to account for the Spitzer/MIPS (\(\SI{70}{\micro \meter}\)) and Herschel/PACS (\(\SI{100}{\micro \meter}\)) far infrared fluxes.

\begin{figure}
\begin{minipage}[c]{0.45\textwidth}
    \includegraphics[width=\textwidth]{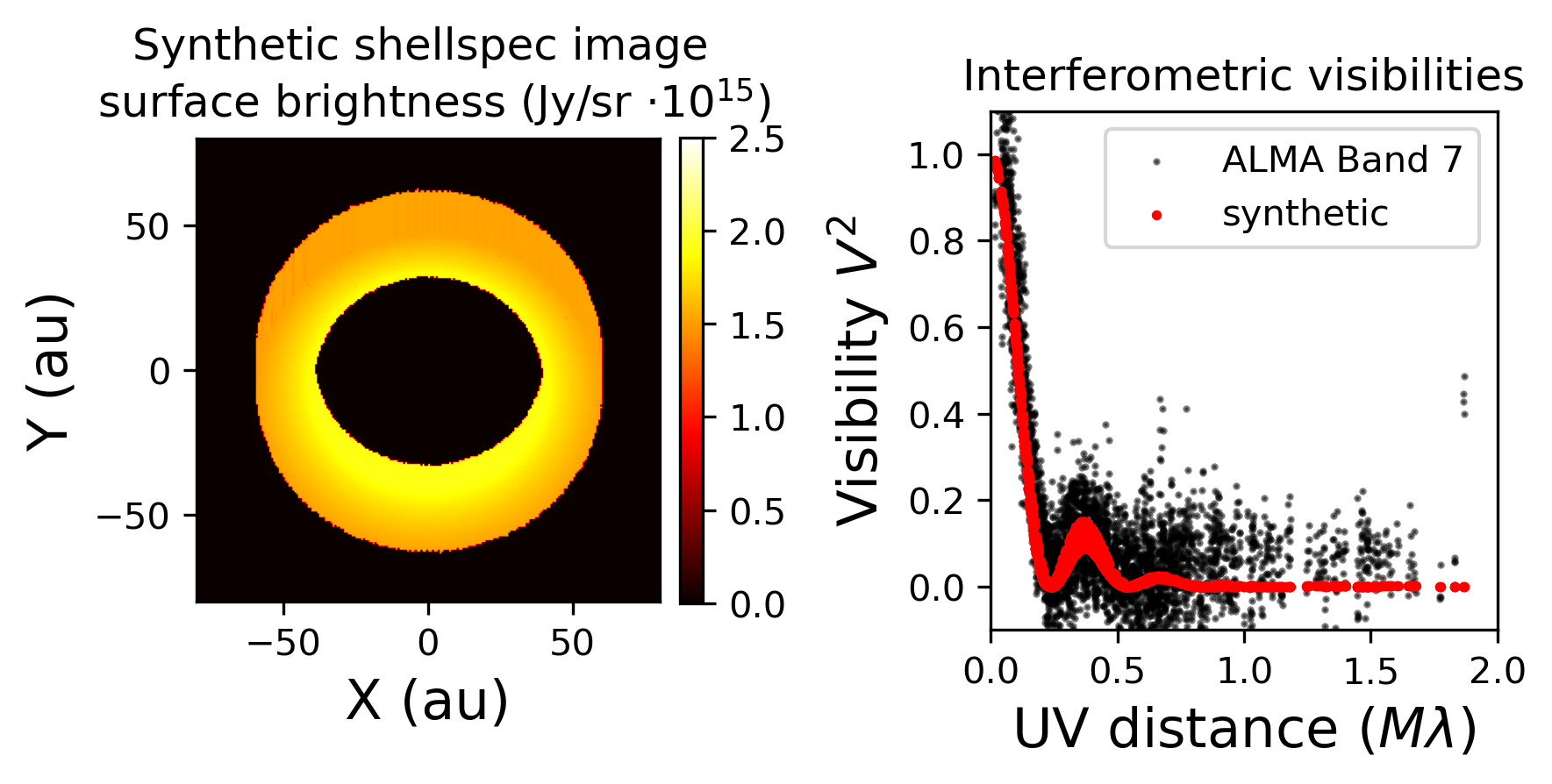}
\end{minipage}\hfill
\begin{minipage}[c]{0.45\textwidth}
    \caption{\textbf{Left:} Synthetic image in the \(\SI{870}{\micro \meter}\) continuum of the best-fit model of the outer disc of DoAr\,44. \textbf{Right:} Fit of the ALMA Band 7 squared visibilities}  \label{fig:outer_model}
\end{minipage}
\end{figure}

\subsection{Intermediate disc model}

A model consisting of the innermost region and the outer disc is enough to account for the spectra, interferometric observables and the SED measurements in the visible to near-IR, as well as the sub-mm and mm regions. However, a deficit in the SED is present in the mid-IR, which has to be compensated for by introducing a warm intermediary region in the 0.2 -- 1\,au range.
We modelled this region using a \texttt{DISC} object, and its parameters can be found in Tab.~\ref{tab:bestfit}. The global density and temperature profiles can also be found in Tab.~\ref{tab:bestfit}.

The temperature at the inner edge reaches \SI{1000}{\kelvin}, and drops to \SI{380}{\kelvin} at the outer edge. \citet{salyk2015detection} detected a large water-vapour luminosity originating from DoAr~44 using the Gemini-North telescope, and deduced the temperature of this water vapour to \(T \approx \SI{450}{\kelvin}\). In our model, these temperatures are reached precisely in this intermediate disc.

It appears that this intermediate disc region has been imaged, but unresolved, by the ODISEA programme in Band~6 \citep{cieza2021ophiuchus} -- see Fig.~\ref{fig:multi}.H.
However, the dimensions of this intermediate disc are not constrained by interferometry or other spatially resolved methods. This means that there exists a large set of acceptable models of this region, extending up to 5 -- 7\,au.

\begin{table}
\caption{
Best-fit parameters of the global model.}
\label{tab:bestfit}
\begin{tabular}{llrrl}
\hline \hline
Parameter & Object & \multicolumn{1}{l}{Best-fit} & \multicolumn{1}{l}{\(\sigma\)} & Unit \\ \hline
\( r_\star \) & \texttt{STAR} & 2.0 & 0.2 & \( R_\odot \) \\
\(T_\star\) & \texttt{STAR} & 4750 & 150 & K \\
\(M_\star\) & \texttt{STAR} & 1.4 & 0.2 & \( M_\odot \) \\ \hline
\( r_{\rm insh} \) & \texttt{SHELL} & 3 & 0.1 & \( R_\odot \) \\
\( r_{\rm outsh} \) & \texttt{SHELL} & 22 & 1 & \( R_\odot \) \\
\( v_{\rm sh} \) & \texttt{SHELL} & 385 & 5 & \si{\kilo \meter \per \second} \\
\( \varrho_{\rm sh} \) & \texttt{SHELL} & {\(1.8\cdot10^{-13}\)} & {\(3\cdot10^{-14}\)} & \( \si{ \gram \per \centi \meter \cubed}\) \\
\( T_{\rm sh} \) & \texttt{SHELL} & 9100 & 200 & K \\
\( e_{\rm velsh} \) & \texttt{SHELL} & 1 & 0.1 & {1} \\ \hline
\( r_{\rm innb1} \) & \texttt{DISC} & 16.5 & 3 & \( R_\odot \) \\ 
\( r_{\rm outnb1} \) & \texttt{DISC} & 35 & 5 & \( R_\odot \) \\
\( \varrho_{\rm nb1} \) & \texttt{DISC} & {\(1.0\cdot10^{-8}\)} & {\(3\cdot10^{-9}\)} & \( \si{ \gram \per \centi \meter \cubed}\) \\
\( T_{\rm nb1} \) & \texttt{DISC} & 1900 & 200 & K \\
\( e_{\rm dennb1} \) & \texttt{DISC} & -0.5 & 0.1 & {1} \\
\( e_{\rm tmpnb1} \) & \texttt{DISC} & -0.7 & 0.1 & {1} \\ \hline
\( r_{\rm innb2} \) & \texttt{DISC} & 43 & 10 & \( R_\odot \) \\
\( r_{\rm outnb2} \) & \texttt{DISC} & 215 & 20 & \( R_\odot \) \\
\( \varrho_{\rm nb2} \) & \texttt{DISC}& {\(6\cdot10^{-10}\)} & {\(1\cdot10^{-10}\)} & \( \si{ \gram \per \centi \meter \cubed}\) \\
\( T_{\rm nb2} \) & \texttt{DISC}& 1000 & 20 & K \\
\( e_{\rm dennb2} \) & \texttt{DISC} & -1.3 & 0.1 & {1} \\
\( e_{\rm tmpnb2} \) & \texttt{DISC} & -0.6 & 0.1 & {1} \\ \hline
\( r_{\rm innb3} \) & \texttt{DISC} & 8600 & 200 & \( R_\odot \) \\
\( r_{\rm outnb3} \) & \texttt{DISC} & 12900 & 200 & \( R_\odot \) \\
\( \varrho_{\rm dstnb3} \) & \texttt{DISC}& \(1\cdot10^{-14}\) & {*} & \( \si{ \gram \per \centi \meter \cubed}\) \\
\( T_{\rm nb3} \) & \texttt{DISC} & 60 & 10 & K \\
\( e_{\rm dennb3} \) & \texttt{DISC} & -2.5 & 0.1 & {1} \\
\( e_{\rm tmpnb3} \) & \texttt{DISC} & -0.8 & 0.1 & {1} \\ \hline
Global model & \(\chi^2_{\rm SED} \)& 7022 & \(N_{\rm SED}\)&  128\\
Accretion region & \(\chi^2_{\rm SPE}\) & 201 & \(N_{\rm SPE} \)& 200\\
Accretion region & \(\chi^2_{\rm V2,VLTI}\) & 8648 & \(N_{\rm IF,VLTI}\)& 512\\
Accretion region & \(\chi^2_{\rm CP}\) & 170 & \(N_{\rm CP}\)& 416\\
Outer disc & \(\chi^2_{\rm V2,ALMA}\) & 18689 & \(N_{\rm V2,ALMA}\) & 3886\\\hline
\end{tabular}
\tablefoot{Model parameters are separated into parameters belonging to the 1. star, 2. shell, 3. innermost disc, 4. intermediary disc, 5. outer disc. (*) The dust density of the outer disc is only given as a lower bound. The \(\chi^2_{\rm SED}\) is computed based on the global model, whereas the other contributions were calculated in the separate components.}
\end{table}

\begin{figure}
\centering
\begin{minipage}[c]{0.45\textwidth}
    \includegraphics[width=\textwidth]{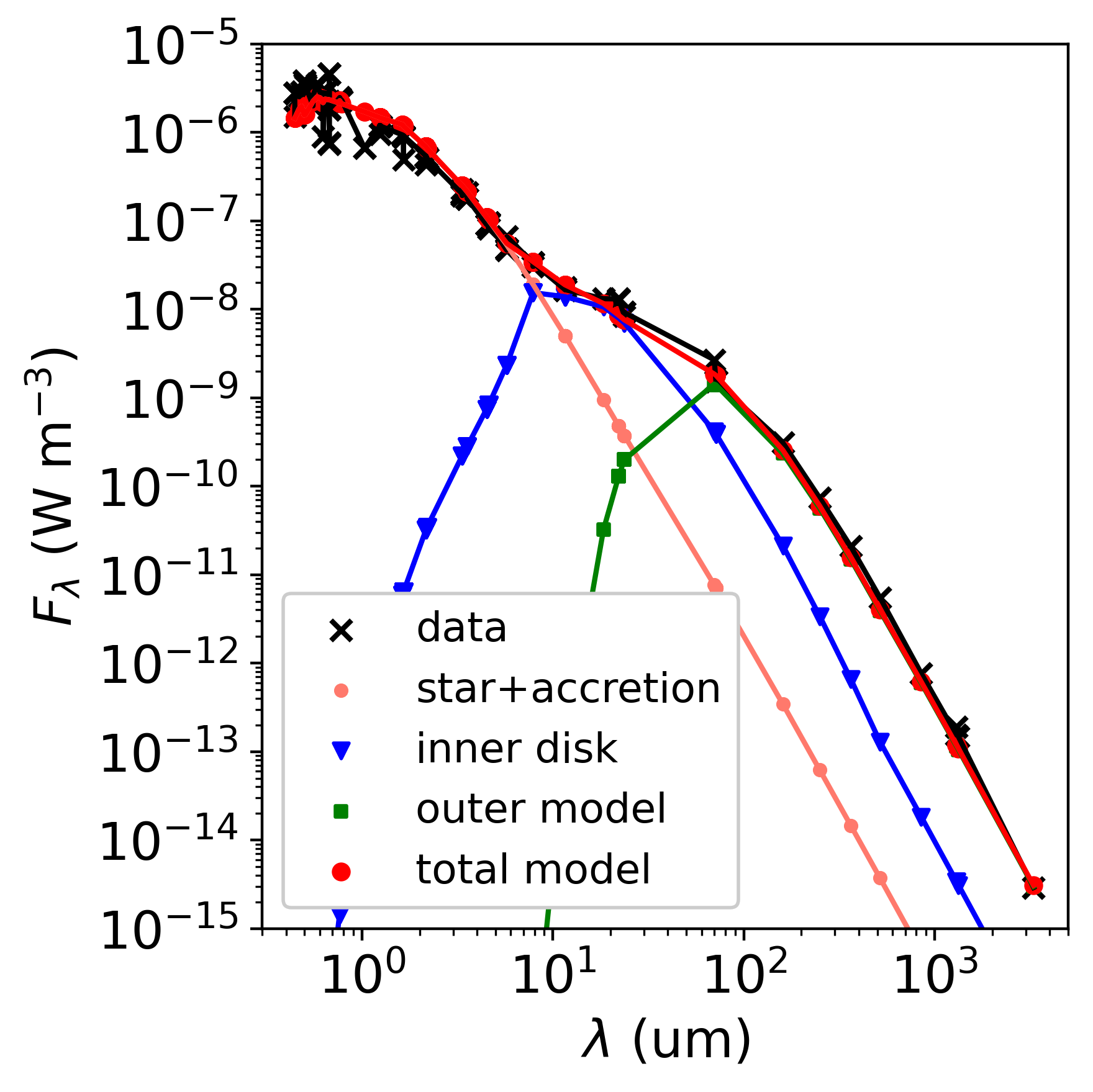}
\end{minipage}\hfill
\begin{minipage}[c]{0.45\textwidth}
    \caption{SED measurements of DoAr\,44 and a synthetic SED, combined out of several components. The central star and accretion region dominate in the ultraviolet and visible range, and the outer disc is dominant in the sub-millimeter and millimeter wavelengths. These two contributions are well constrained from VLTI and ALMA interferometry. However, if only these two components are considered, a large deficit is present in the mid-infrared range. Therefore, an intermediate region must be present. The error-bars are not visible, as the vertical axis spans ten magnitudes.}  \label{fig:SED_comb}
\end{minipage}
\end{figure}

\begin{figure}
    \centering
    \includegraphics[width=\linewidth]{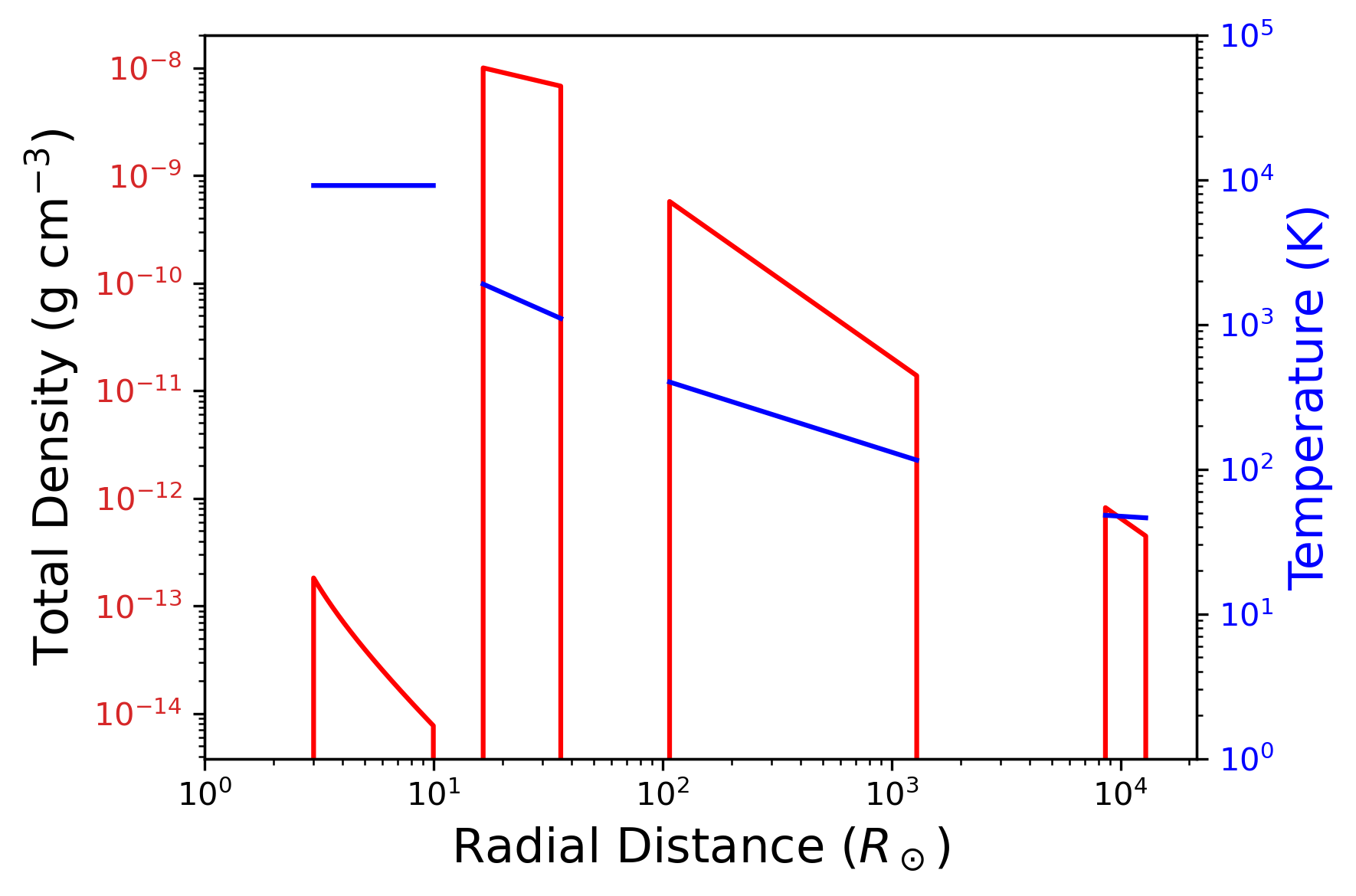}
    \caption{
    The total (gas + dust) density
    and temperature profiles of DoAr\,44, with the assumed metallicity of $Z = 0.0122$.
    The model is composed of four segments (from left to right): 1. A hot, optically thin shell, where H\(\alpha\) emission originates. 2. An innermost disc, which presumably feeds the accretion shell. 3. An intermediate disc, necessary to explain the SED in the mid-IR region. 4. A cold outer disc, visible with the ALMA. Between the intermediate and the outer disc, a large \({\approx}30\)\,au cavity is present.}
    \label{fig:rho_and_T}
\end{figure}

\section{Discussion}

We shall now discuss the broader implications of our global model of DoAr~44
for the disc geometry, condensation of solids, and planet formation.

\subsection{What is the geometry of the accretion region?}

As discussed in Sect.~\ref{Halpha},
the detailed geometry of the inner accretion region is variable on the scale of days, and is most likely connected to magnetic funnel flows, described in \citet{bessolaz2008accretion, takasao2018three, bouvier2020investigating}. We have modelled the inner optically thin matter responsible for the averaged H\(\alpha\) emission using axially symmetric conical shapes, but optimising always resulted in large opening angles, or a spherically symmetric shell of gas. 

Around this optically thin shell of gas, we include an optically thick innermost disc to explain the VLTI/GRAVITY visibilities. In our model, the innermost disc appears to be misaligned by about \((18.5 \pm 4)\degree\) with respect to the outer disc. The topic of misaligned protoplanetary discs has gained significant attention in recent years, as these discs could produce exoplanets on highly inclined orbits, which are now readily observed using the Rossiter-McLaughlin effect \citep{zak2024}.

A variety of mechanisms to produce a misaligned disc have been identified, e.g., compact stellar binaries \citep{nixon2013}, using an outside perturbing companion \citet{dogan2015}, stellar fly-bys \citep{nealon2019} or oblique infall \citep{kuffmeier2021}. 
In our case, we can rule out the presence of a compact binary within the inner disc from the VLTI/GRAVITY observations, no infall is currently observed with ALMA, and no planetary companions have been confirmed, so the origin of this misalignment remains uncertain.

\subsection{Does the observed ring coincide with a condensation line of a volatile, or must it be caused by an exoplanet?}

The overall temperature profile in our model predicts around 50 -- 60\,K
around the distance of 48\,au, in order to account for far-infrared continuum radiation.
This relatively high temperature of layers emitting the respective radiation,
e.g., compared to \citet{leiendecker2022dust}, 
prevents condensation of certain solids \citep{martin2014thermal}.

The scientific literature on solar system comets, which originate from these cold outer regions, is a good guide with respect to the expected chemical composition. The major volatiles found in comets are H$_2$O, CO, CO$_2$, CH$_4$, and NH$_3$, with water ice being typically the most abundant \citep{mumma2011chemical}. The comet's composition is often given with respect to the water ice content. CO can contribute up to 30\,\% of the mass of the water ice, CO$_2$ up to 10\,\%, and the other chemicals generally contribute less than 1\,\% \citep{1998A&A...330..375G}. Different ratios were obtained for example by Rosetta orbiting 67P/Churyumov-Gerasimenko, where CO$_2$ formed 7.5\,\% and CO 3\,\% \citep{10.1093/mnras/stad3005}. These ratios might be lower than what we expect in the outer disc, as 67P/Churyumov-Gerasimenko is a periodic comet, which effectively loses highly volatile ices during every perihelion passage.

Water ice does not seem like a good candidate for the condensation line at this distance. Given the temperature profile from our models, the water ice line occurs somewhere between 4--6\,au, somewhat similar to the \(\approx 3\)\,au water ice line distance in our early solar system. 

The condensation of carbon monoxide is not responsible for the continuum emission at this distance, because its condensation temperature should be 23 to 28\,K
\citep{Zhang_2013}.

However, we cannot exclude the possibility that CO is condensed in the mid-plane,
which could be colder than 50 to 60\,K,
but the mid-plane is inaccessible due to optical thickness.

Possibly, carbon dioxide could be transitioning from the gaseous to the solid state at the peak emission radius in DoAr\,44. However, we must proceed with caution: solid CO$_2$ is associated with mid-IR bands, which would be in the range observed by the Spitzer/IRS instrument. These CO$_2$ bands around \(\SI{15}{\micro \meter}\) are often observed around YSOs \citep{cook2011thermal}. Due to the complicated nature of the silicate spectral features that overlap with the possibly present solid CO$_2$ feature, we have decided to refrain from further interpretations.

The possibility of other solids, forming the ring in DoAr\,44, seems improbable due to the total dust mass in the outer disc. There is probably more than \(100\, M_{\rm Earth}\) of dust in the outer disc, which means it cannot be composed of some rare chemical.

\subsection{The planet formation mass budget}

According to the outer disc model, depending mainly on the representative particle size,
ranging from fine \(\SI{1e-2}{\micro \meter}\) to coarser \(\SI{1e2}{\micro \meter}\),
we obtain the lower limit of dust mass,
$M_{\rm dust} > 100\, M_{\rm Earth}$.

Coincidentally, such a value is similar to what is commonly presumed for our own solar system ( \({\approx}130\,M_{\rm Earth}\), \citet{brovz2021early}),
but the latter is for the whole dust disc, not a narrow ring.
This leads us to think that the ring structure at 48\,au, seen in the ALMA data, might very well be an analogue of the Kuiper belt in our own solar system, which currently extends from about 30\,au to about 55\,au, and presumably had around 15 -- 20 $M_{\rm Earth}$ in planetesimals \citep{Nesvorny_2018ARA&A..56..137N}.

Extrapolating the outer disc surface density to the inner disc, using the \citet{hayashi_mmsn} surface density exponent of \(\Sigma \propto r^{-1.5}\), we obtain a hypothetical amount of solids in the cavity from 6 to 40\,au around \(225\, M_{\rm Earth}\); using a shallower exponent of $-1.0$, the result changes to \({\approx}150\,M_{\rm Earth}\)
(see also Fig.~\ref{fig:rho_and_T}).
This suggests that former, "Class~I" disc of DoAr\,44 was more massive
than in our solar system and that the cavity should contain more,
or more massive planets.

\section{Conclusions}

In this work, we present the first global radiative-transfer model of the protoplanetary disc DoAr\,44, based on VLT/UVES and VLT/X-shooter spectra, VLTI/GRAVITY squared visibilities, triple product amplitudes and closure phases, ALMA interferometry, and a variety of spectral energy distribution measurements, spanning from the near ultraviolet to millimetre wavelengths. From this model, we have concluded the following:

\begin{enumerate}
    \item The geometry of the inner accretion region can be described using a shell of gas, heated to about 9\,000\,K, surrounding the young star. This gas is moving at speeds up to \(\SI{385}{\kilo \meter\,\second^{-1}}\). This shell is surrounded by an innermost disc (\(16\,R_\odot\) to  \(35\,R_\odot\), i.e., 0.07 to 0.16\,au), which is responsible for the VLTI/GRAVITY K-band visibilities.
    
    \item The accretion rate is already low, \(\dot{M} = (6.1 \pm 0.2)\cdot10^{-8}\,M_\odot\, {\rm yr}^{-1}\), but not negligible.
    
    \item The continuum dust emission in the outer disc extends from about (\(36\pm1\))\,au to about \((60\pm4)\)\,au. A large part of the emission is localized in a narrow ring centered at (\(48\pm1\))\,au. The temperature in this region is about 60\,K. The observations are not compatible with a monotonic density profile, and a large central cavity must be present. This cavity was likely cleared by forming ice/gas planets.
    
    \item The ring at (\(48\pm1\))\,au does not correspond to the condensation line of water ice or carbon monoxide due to high temperature (50 -- 60\,K). The only possible condensation line that could cause such a ring is carbon dioxide.
    
    \item The dust mass present in the outer disc is more than \(100\,M_{\rm Earth}\), according to the integral of the surface density profile.
     
    \item An intermediate disc, for the moment unresolved by ALMA, extending from 0.2 to 1 au or beyond, must be present, according to the SED constraints. From this region, the observed water vapour emission probably originates.
\end{enumerate}

\begin{acknowledgements}
We would like to thank the anonymous referee for their time and care, which helped improve this work. We would also like to thank S. Casassus for help with the interpretation of some ALMA measurements.

This work has been supported by the Czech Science Foundation (grant no. 25-16507S).

This paper makes use of the following ALMA data sets: ADS/JAO.ALMA\#2019.1.00532.S, ADS/JAO.ALMA\#2019.1.01111.S. ALMA is a partnership of ESO (representing its member states), NSF (USA) and NINS (Japan), together with NRC (Canada), NSTC and ASIAA (Taiwan), and KASI (Republic of Korea), in cooperation with the Republic of Chile. The Joint ALMA Observatory is operated by ESO, AUI/NRAO and NAOJ.

Based on data obtained from the ESO Science Archive Facility with DOI(s):
\href{https://doi.eso.org/10.18727/archive/21}{https://doi.eso.org/10.18727/archive/21},
\href{https://doi.eso.org/10.18727/archive/50}{https://doi.eso.org/10.18727/archive/50},
\href{https://doi.eso.org/10.18727/archive/71}{https://doi.eso.org/10.18727/archive/71}.

Based on observations collected at the European Organisation for Astronomical Research in the Southern Hemisphere under ESO programme(s) 69.C-0481, 093.C-0658, 085.C-0764.

This work has made use of data from the European Space Agency (ESA) mission
{\it Gaia} (\url{https://www.cosmos.esa.int/gaia}), processed by the {\it Gaia}
Data Processing and Analysis Consortium (DPAC,
\url{https://www.cosmos.esa.int/web/gaia/dpac/consortium}). Funding for the DPAC
has been provided by national institutions, in particular the institutions
participating in the {\it Gaia} Multilateral Agreement.   
\end{acknowledgements}

\bibliographystyle{aa} % style aa.bst
\bibliography{z_bibliography_cleaned_updated}

\begin{appendix}
\clearpage
\section{Additional figures}

\begin{figure}[h]
    \centering
    \includegraphics[width=\linewidth]{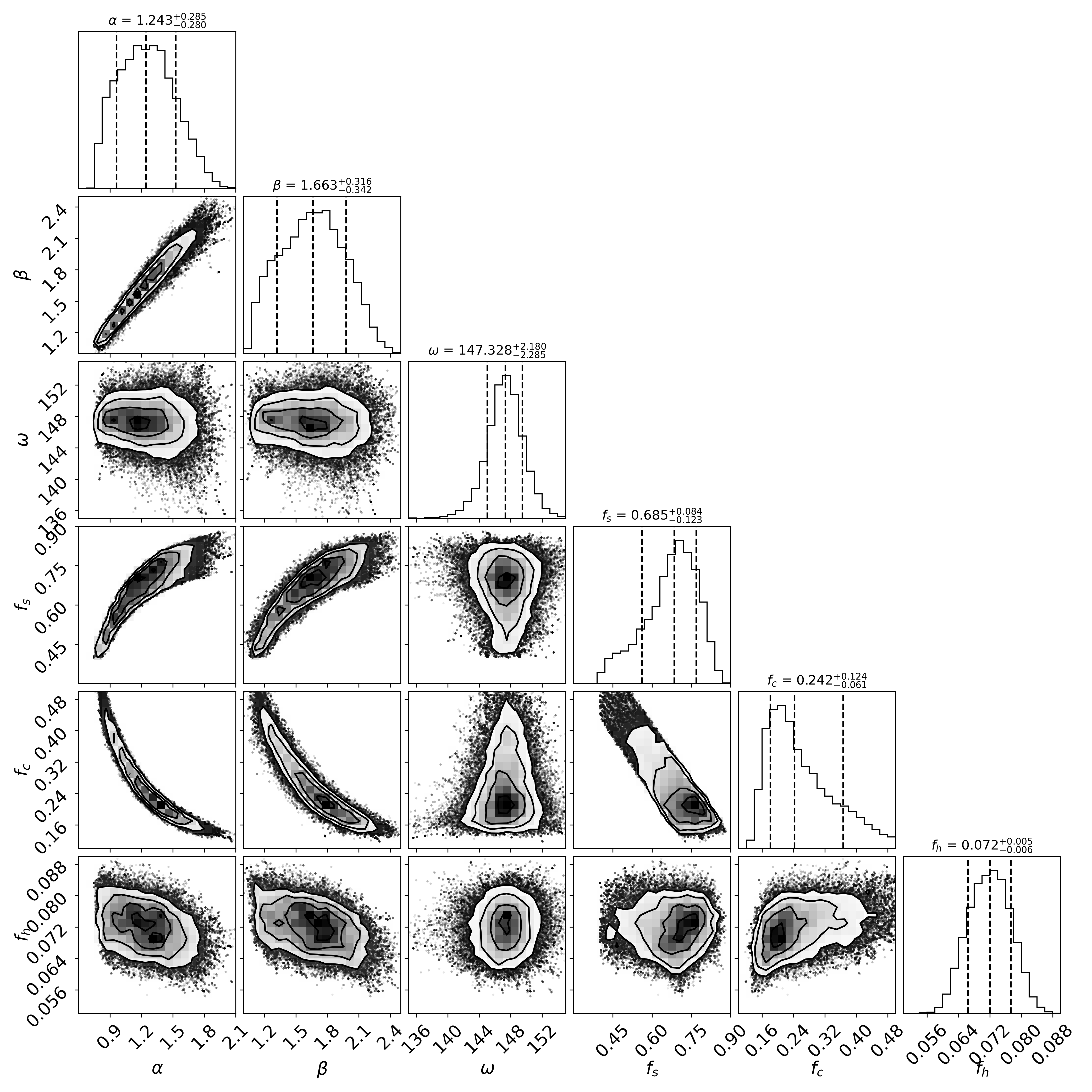}
    \caption{For DoAr\,44, the joint posterior and marginalized probability distributions for the homogeneous ellipse model, based on VLTI/GRAVITY measurements, are presented. These parameters are \(\alpha\) and \(\beta\), describing the ellipse's semi-major and semi-minor axes in milli-arcseconds, \(\omega\) describing the position angle of the ellipse, and \(f_{\rm s}, f_{\rm c}, f_{\rm h}\) giving the fractional contributions of the star, circumstellar matter and a diffuse, ``always resolved" halo. Although some parameters exhibit strong correlations, this does not affect the usefulness of the calculations. The size of the ellipse (at 146\,pc) is 0.2 \(\times\) 0.24\,au, and the contribution of the always-resolved halo is approximately \(7\%\). The parameter uncertainties appear underestimated, likely due to the simplicity of the model.
}
    \label{fig:cornerplot_mod2}
\end{figure}
\begin{figure}[h]
    \centering
    \includegraphics[width=\linewidth]{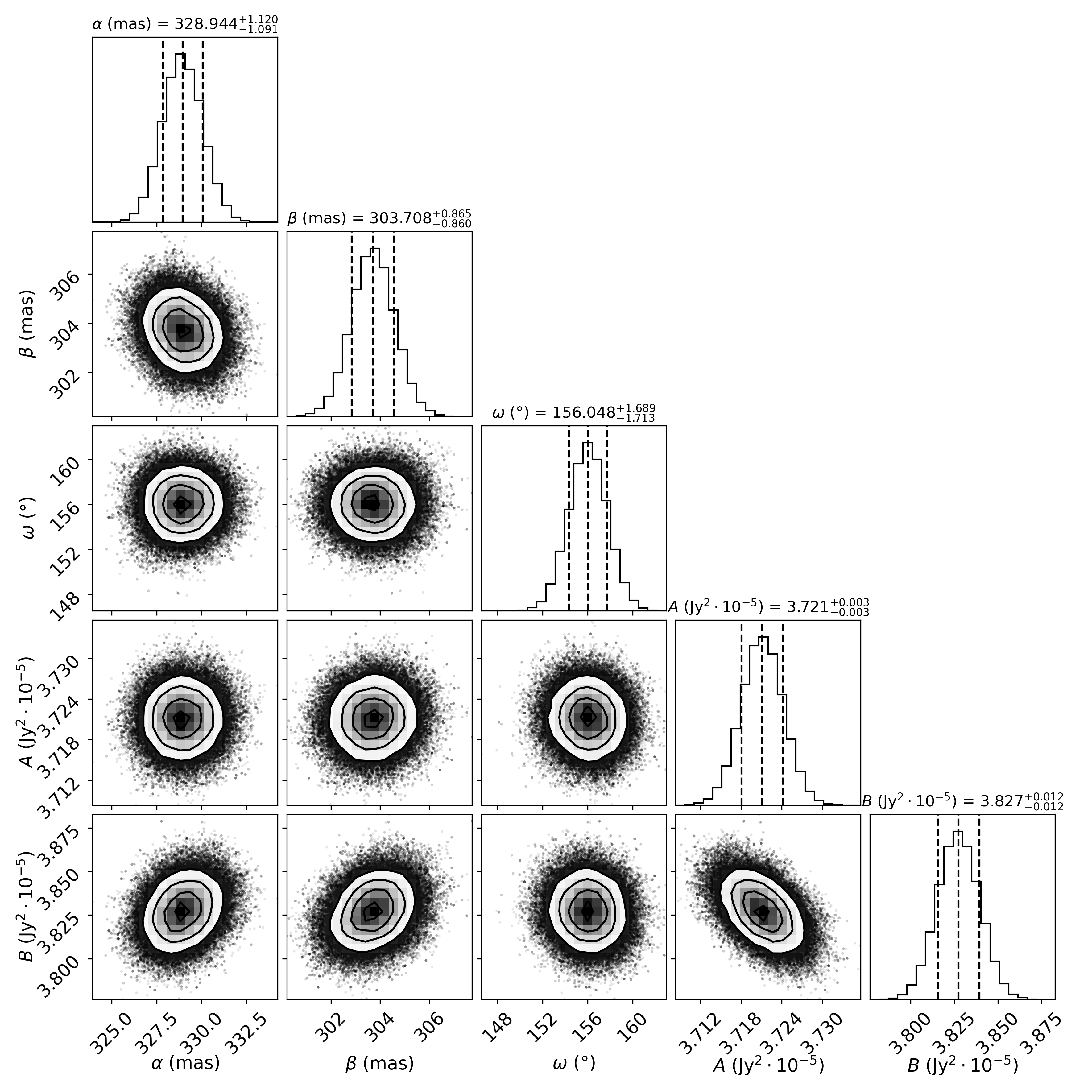}
    \caption{For DoAr\,44, the joint posterior and marginalized probability distributions for the parameters of the \(\delta\)-ring model describing the ALMA Band 7 data are presented. These parameters are \(\alpha\) and \(\beta\), describing the dimensions of the ellipse in milli-arcseconds, \(\omega\), giving the position angle in degrees, and \(A\) and \(B\), representing the fractional contribution of the unresolved and always-resolved emission, respectively. 
}
    \label{fig:Delta_ring_mcmc}
\end{figure}
\clearpage
\begin{figure}[b]
\centering
    \onecolumn\includegraphics[width=\textwidth]{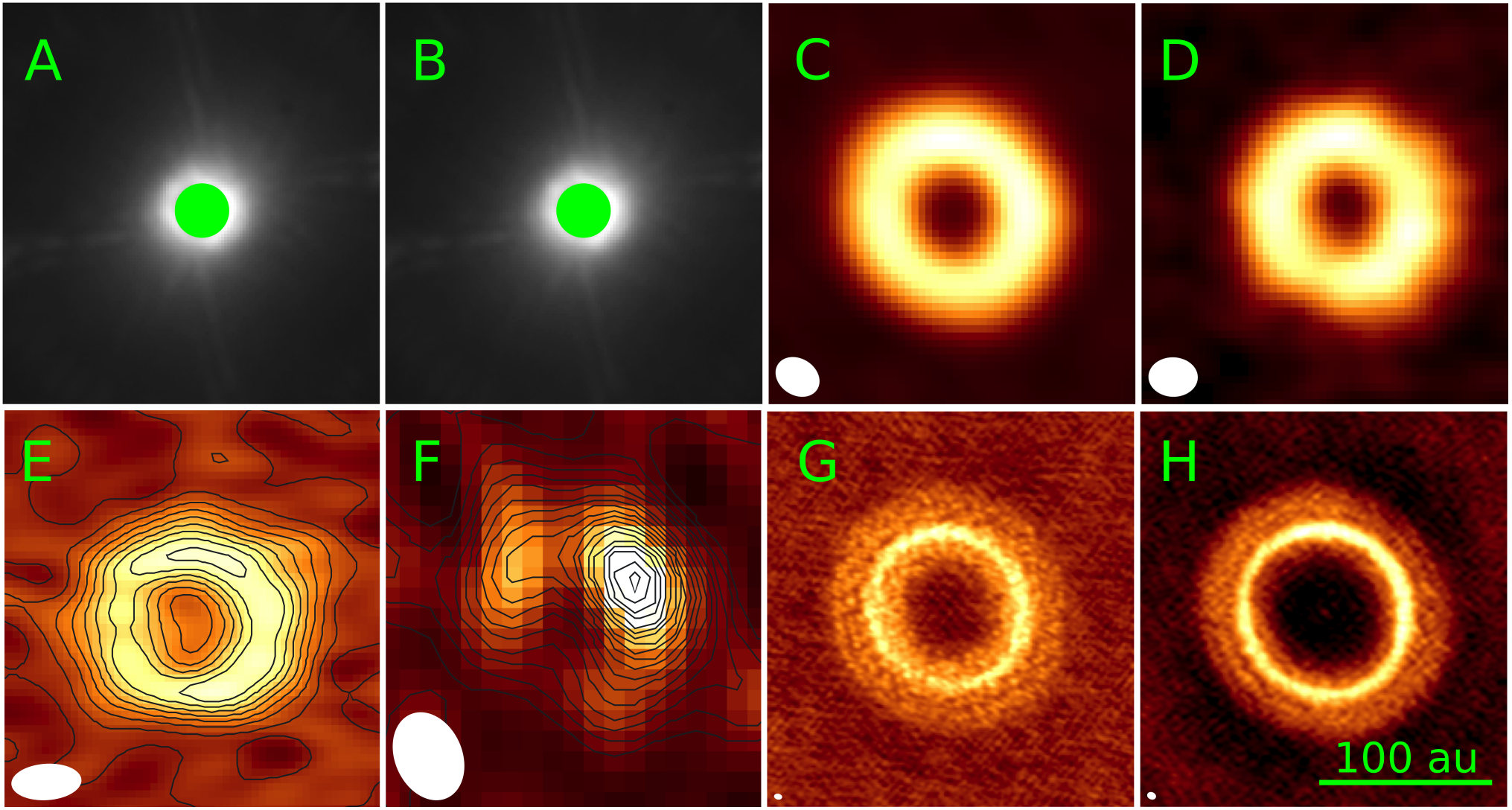}
    \caption[Available spatially resolved-images of the source DoAr\,44 (SPHERE, SMA, ALMA)]{A comparison of available spatially resolved observations of DoAr\,44. \textbf{A}: VLT/SPHERE/IRDIS Channel 1 \citep{avenhaus2018disks}. \textbf{B}: VLT/SPHERE/IRDIS Channel 2 \citep{avenhaus2018disks}.
    \textbf{C}: ALMA Band 3. Angular resolution: 0.251''. Project code: 2019.1.01111.S. 
    \textbf{D}: ALMA Band 7. 0.168'' \citep{arce2023radio}.   \textbf{E}: ALMA Band 8.   0.154'' \citep{2023MNRAS.526.1545C}.  \textbf{F}: SMA at 346 GHz. 0.41'' \citep{2016A&A...585A..58V}. \textbf{G}: ALMA Band 7. 0.023'' \citep{arce2023radio}. \textbf{H}: ALMA Band~6. 0.026'' \citep{cieza2021ophiuchus}. The SPHERE images are in the H band, and correspond to perpendicular linear polarizations. The green circle on the SPHERE images is the 0.1'' radial extent of the coronagraph. Contours are plotted over the low-resolution interferometric images. The synthesized beam, corresponding to the angular resolution, is shown as a white ellipse (very small for plots G and H). The scale is consistent on all plots. These ALMA observations were also used to constrain the model of the SED.}\label{fig:multi}
\end{figure}

\end{appendix}

\end{document}